\documentclass[aps,prd,showkeys,showpacs,twocolumn,superscriptaddress]{revtex4}

\usepackage{graphics}
\usepackage{amsfonts}
\usepackage{amsmath}
\usepackage{latexsym}
\usepackage{amssymb}
\usepackage[mathcal]{euscript}
\usepackage{eurosym}
\usepackage{rotating}

\newcommand{\ket}[1]{| #1 \rangle}
\newcommand{\bra}[1]{\langle #1 |}

\newcommand{\trace}{\textrm{Tr}}

\newcommand{\herm}{\textrm{Herm}^+_{d}(\mathbb{C})}

\newcommand{\europ}{\textup{\emph{\geneuro}}}

\newcommand{\widecheck}[1]{\overset{\begin{turn}{180}$\widehat{}$\end{turn}}{#1}}
\newcommand{\wheurop}{\widehat{\europ}}

\newcommand{\whdoll}{\widehat{\$}}
\newcommand{\wcdoll}{\widecheck{\$}}
\newcommand{\lonewh}{\widehat{\phantom{U}}}
\newcommand{\lonewc}{\widecheck{\phantom{U}}}
\newcommand{\whx}[1]{\widehat{#1}}
\newtheorem{Def}{Definition}
\newtheorem{Rk}{Remark}
\newtheorem{Iso}{Isomorphism}
\newtheorem{Th}{Theorem}
\newtheorem{Proposition}{Proposition}
\newtheorem{Pro}{Property}
\newtheorem{Lem}{Lemma}
\newtheorem{Cor}{Corollary}

\begin{document}

\title{\begin{center}
On quantum operations as quantum states
\end{center}
}

\author{Pablo Arrighi}
\email{pja35@cam.ac.uk} \affiliation{Computer Laboratory,
University of Cambridge, 15 JJ Thomson Avenue, Cambridge CB3 0FD,
U.K. }
\author{Christophe Patricot \setcounter{footnote}{5}}
\email{cep29@cam.ac.uk} \affiliation{ DAMTP,
University of Cambridge, Centre for Mathematical Sciences,\\
Wilberforce Road, Cambridge CB3 0WA, U.K.}



\begin{abstract}
We formalize the correspondence between quantum
states and quantum operations isometrically, and harness its consequences. This
correspondence was already implicit in the various proofs of the
operator sum representation of Completely Positive-preserving
linear maps; we go further and show that all of the important
theorems concerning quantum operations can be derived directly from
those concerning quantum states. As we do so the
discussion first provides an elegant and original review of the
main features of quantum operations. Next (in the second half of
the paper) we find more results stemming from our formulation of the
correspondence. Thus we provide a factorizability condition for
quantum operations, 
and give two novel Schmidt-type decompositions of bipartite pure
states. By translating the composition law of quantum operations, we
define a group structure upon the set of totally
entangled states. The question
whether the correspondence is merely mathematical or can be given
a physical interpretation is addressed throughout the text:  we
provide formulae which suggest quantum states inherently define a
quantum operation between two of their subsystems, and which turn out
to have applications in quantum cryptography. 
\end{abstract}

\maketitle 
This article is concerned with the properties of
positive matrices (quantum states) and the linear maps between
these, i.e. Positive-preserving linear maps and  Completely
Positive-preserving linear maps (quantum operations), as provided
by the density matrix formalism of finite dimensional quantum
theory. The analysis we carry out is formal and mathematical, and
although it focuses on some quantum
information theoretical issues,  it should have applications in other
domains. The driving line of the article is in its method:
formalizing and exploiting systematically an isomorphism from
hermitian matrices to Hermitian-preserving linear maps and quantum
states to quantum operations. To our knowledge, this isomorphism was 
first used by Sudarshan et Al.  \cite{Su} in the quantum theoretical context, 
and was later popularized by Jamiolkowski \cite{Jamio}, and Choi \cite{Choi}. 
The operator sum representation theorem has been independently derived by Kraus 
\cite{Kraus1} (see also \cite{Kraus}) -- with a proof valid in infinite dimensions. 
Our investigation shows that the isomorphism between states and operations
has a much wider range of implications, whether to simplify the proofs of
well-known results or to point out novel properties, both technical
and geometrical. The presentation is rigorous and self-contained, we give all the
necessary background for someone to enter the subject.

In section \ref{setting}, after setting our conventions, we relate
vectors to matrices, and matrices to superoperators, the idea
being to map an $mn\times mn$ matrix to a linear operator from
$n\times n$ matrices to $m\times m$ matrices. These isomorphisms
are often viewed pragmatically as rearrangements of the
coordinates of vectors or matrices, but we formalize them
more abstractly as norm-preserving bijections between tensor product
spaces. We derive original formulae relating to these
isomorphisms which we use throughout the article. One of them will
simplifiy those numerous mathematical problems in quantum cryptography which
require a careful optimization of the fidelities induced by a quantum
operation. This formal
setting leads in subsection \ref{the correspondence} to the
state-operator equivalence, inherently present in the works many,
but rarely exploited as such: non-normalized quantum states of an
$mn$-dimensional system are equivalent to quantum operations from
an $n$-dimensional system to an $m$-dimensional one. \\
We use this correspondence in subsection \ref{properties} to
rederive all the main properties of quantum operations from those
of quantum states: the operator sum decomposition and its unitary
degree of freedom stem from the spectral decomposition and
Hughston-Josza-Wooters theorems; the factorizability of quantum
operations up to a trace-out corresponds to the purification of
quantum states; and the polar decomposition of matrices is
equivalent to the Schmidt decomposition of pure states. Next, in
subsection \ref{new properties}, we consider properties of states
(or operations) whose translation in terms of operators (or states)
was unknown to us previously. Mainly we give a factorizability
condition for quantum operations, i.e. a criteria for an operator
to be single operator in the operator sum representation; and we find two
original triangular decompositions of pure states of a bipartite system.
Throughout the section the normalization of density matrices is
unimportant. Yet for completeness the reader is reminded of the
well known Trace-preserving conditions in subsection \ref{trace
preservation subsection} (both in terms of states and
operators). Moreover we highlight the fact that maximally entangled
pure states of  a bipartite system go
hand in hand with isometric maps from one subsystem to the other
(unitary maps in case both systems have the same dimension).
Choi's extremal Trace-preserving condition is also presented and
recasted in terms of the rank of an easily constructed matrix.\\
Section \ref{geometrical structure} is devoted to geometrical
structures of quantum states. We exploit the composition law on
Completely Positive-preserving maps to define a semi-group
structure on the states of $n^2$-dimensional quantum systems, and
show that the subset of totally entangled pure states is
isomorphic to the group of invertible $n\times n$ matrices defined
up to phase (with maximally entangled pure states corresponding to
unitary transforms as in \cite{Landau}). These group
isomorphisms have profound structural meaning, and are useful in
finding nice coordinate charts on such spaces. We also give an
exotic composition law on operators stemming from the Schur
product on states. In subsection \ref{duality} we make use of the
dual mapping between states and positive functionals, and readily
show that the space of Positive-preserving maps is dual to that of
separable states of a bipartite system. This yields a simple
result which is in fact equivalent to Peres' separability
criterion. More generally the notion of duality seems to help
provide possible physical interpretations of the state-operator
correspondence formulae, notably as we show that the effect of any
quantum operation can be viewed as the trace out of a particular
local single operation on its corresponding state.\\
We conclude in section \ref{conclusion} and give a table summarizing the main
results.

\section{The setting}
\label{setting} We denote by $M_d(\mathbb{C})$ the set of $d\times
d$ matrices of complex numbers, and by
$\textrm{Herm}_d(\mathbb{C})$ its hermitian subset. Amongst the
latter we will denote by $\textrm{Herm}_d^{+}(\mathbb{C})$ the set
of positive matrices, and also refer to it as the set of
(non-normalized) states of a $d$-dimensional quantum system. An
important subset of $\textrm{Herm}_{mn}^+(\mathbb{C})$ is the set
of \emph{separable states}, i.e. those which can be written in the
form
\begin{equation*}
\rho= \sum_{x}\lambda_x \rho_1^x \otimes \rho_2^x
\end{equation*}
where $\lambda_x\geq 0$ and the $\rho_1^x$ and $\rho_2^x$ belong
to $\textrm{Herm}_m^+(\mathbb{C})$ and
$\textrm{Herm}_n^+(\mathbb{C}^)$ respectively. Later we shall
denote this set by $\textrm{Herm}_{mn}^S(\mathbb{C})$.\\
Throughout the dagger operation $^{\dagger}$ will be somewhat
overloaded, in a manner which has now become quite standard: as
usual a ket $A=\sum A_{i}|i\rangle$ will be taken into a bra
$A^{\dagger}=\sum A^*_{i}\langle i|$, while a matrix $\hat{A}=\sum
A_{ij}|i\rangle\langle j|$ will be mapped into its conjugate
transpose $\hat{A}^{\dagger}=\sum A^*_{ij}|j\rangle\langle i|$. In
other words, $^{\dagger}$ takes kets into bras using  the
canonical complex scalar product for vectors, i.e.
$B^{\dagger}\equiv [A\mapsto (B,A)=\sum B^*_i A_i \equiv
B^{\dagger}A]$, but for linear maps on vectors it denotes the
usual adjoint operation defined with respect to the same scalar
product. We also make frequent use of the conjugation operation
$^*$ which is defined in the canonical basis to take kets $A=\sum
A_i\ket{i}$ into $A^*=\sum A^*_i \ket{i}$, and similarly on bras.
Linearity will refer to complex linearity.

\begin{Def}
A linear map $\Omega : M_m(\mathbb{C}) \rightarrow
M_n(\mathbb{C})$ is Hermitian-preserving if and only if for all
$\rho$ in $\textrm{Herm}_m(\mathbb{C})$, $\Omega(\rho)$ belongs to
$\textrm{Herm}_n(\mathbb{C})$.
\end{Def}
The following is a well-known fact:
\begin{Rk}
\label{Hermextends} If $\Omega: M_m(\mathbb{C}) \rightarrow
M_n(\mathbb{C})$ is a Hermitian-preserving linear map, then so is
$\Omega\otimes Id_r$.
\end{Rk}
\emph{Proof.} Let us denote by $\{\tau_i\}$ and $\{\tau_j\}$ two
sets of hermitian matrices forming a basis of
$\textrm{Herm}_m(\mathbb{C})$ and $\textrm{Herm}_r(\mathbb{C})$
respectively, considered as a real vector spaces.
$\{\tau_i\otimes\tau_j\}$ forms a basis for
$\textrm{Herm}_{mr}(\mathbb{C})$. Now consider
$Z\in\textrm{Herm}_{mr}(\mathbb{C})$, so that $Z=\sum_{ij}
z_{ij}\tau_i\otimes\tau_j$ with $z_{ij}\in \mathbb{R}$. We then
have
\begin{align*}
(\Omega\otimes Id_r)Z&=\sum_{ij}
z_{ij}\Omega(\tau_i)\otimes\tau_j \\
&=\sum_{ij} z_{ij}\Omega(\tau_i)^{\dagger}\otimes\tau^{\dagger}_j
=
\sum_{ij}\big(z_{ij}\Omega(\tau_i)\otimes \tau_j \big)^{\dagger} \\
&=((\Omega\otimes Id_r)Z)^{\dagger}\quad\quad\Box
\end{align*}
\begin{Def}
A linear map $\Omega : M_m(\mathbb{C}) \rightarrow
M_n(\mathbb{C})$ is Positive-preserving if and only if for all
$\rho$ in $\textrm{Herm}_m^+(\mathbb{C})$, $\Omega(\rho)$ belongs
to $\textrm{Herm}_n^+(\mathbb{C})$.
\end{Def}
A Positive-preserving map is necesseraly Hermitian-preserving
since any hermitian matrix can be expressed as the difference of
two positive matrices. Note also that having $\Omega:
M_m(\mathbb{C}) \rightarrow M_n(\mathbb{C})$ a Positive-preserving
linear map does not imply
that $\Omega\otimes Id_r$ is also Positive-preserving. \\
\emph{Example.} The map
\begin{align*}
^{t}:\textrm{Herm}_2^+(\mathbb{C})&\rightarrow\textrm{Herm}_2^+(\mathbb{C})\\
\rho&\mapsto\rho^{t}
\end{align*}
is clearly Positive-preserving, but $(^{t}\otimes Id_2)$ is not:
 indeed let
$|\beta\rangle=|00\rangle+|11\rangle,\;|\gamma\rangle=|01\rangle+|10\rangle\;\textrm{and}\;|\delta\rangle=|01\rangle-|10\rangle$
an orthogonal basis of $\mathbb{C}^2 \otimes \mathbb{C}^2$.
\begin{align*}
(^{t}\otimes Id_2)(|\beta\rangle\langle\beta|)&=|00\rangle\langle
00|+|10\rangle\langle 01|+|01\rangle\langle 10|+|11\rangle\langle
11|\\
&=|00\rangle\langle 00|+|11\rangle\langle
11|+|\gamma\rangle\langle\gamma|-|\delta\rangle\langle\delta|
\end{align*}
which is not positive since $\bra{\delta}(^{t}\otimes
Id_2)(|\beta\rangle\langle\beta|)\ket{\delta}<0$.
\begin{Def}
A linear map $\Omega : M_m(\mathbb{C}) \rightarrow
M_n(\mathbb{C})$ is Completely Positive-preserving if and only if
for all $r$ and for all $\rho$ in
$\textrm{Herm}_{mr}^+(\mathbb{C})$, $(\Omega\otimes Id_r)(\rho)$
belongs to $\textrm{Herm}_{nr}^+(\mathbb{C})$.
\end{Def}

\subsection{Isomorphisms}

Next we relate vectors of $\mathbb{C}^m\otimes \mathbb{C}^n$ to
endomorphisms from $\mathbb{C}^n$ to $\mathbb{C}^m$. The tensor
split of $\mathbb{C}^{mn}$ into $\mathbb{C}^m\otimes \mathbb{C}^n$
is considered fixed, as will be all tensor splits throughout the
article unless specified otherwise. (Notions of entanglement will
refer to a particular tensor product of spaces, given a priori.)
Let $\{\ket{i}\}$ and $\{\ket{j}\}$ be orthonormal basis of
$\mathbb{C}^m$ and $\mathbb{C}^n$ respectively, which we will
refer to as canonical.
\begin{Iso}
\label{isomorphism1} The following linear map
\begin{align*}
\hat{}\;:\mathbb{C}^m\otimes \mathbb{C}^n &\rightarrow
End(\mathbb{C}^n \rightarrow \mathbb{C}^m)\\
A&\mapsto\hat{A}\\
\sum_{ij} A_{ij}\ket{i}\ket{j}&\mapsto \sum_{ij}
A_{ij}\ket{i}\bra{j}
\end{align*}
where $i=1,\ldots,m$ and $j=1,\ldots,n$, is an isomorphism taking
vectors $A$ into $m\times n$ matrices $\hat{A}$. It is isometric in the
sense that:
\begin{equation}
\forall A,B \in \mathbb{C}^m\otimes \mathbb{C}^n, \quad  B^{\dagger}A = \trace(\hat{B}^{\dagger}\hat{A})\label{isometry1}
\end{equation}
\end{Iso}
\emph{Proof.} This is trivial, but note that the definition of this
isomorphism is basis dependent. $\quad \Box$ \\
Following a very convenient notation introduced by
Sudarshan\cite{Shu} we will often use a semicolon `;' to separate
output indices (on the left) from input indices (on the right),
together with the repeated indices summation convention. For
instance the matrix $\hat{A}:\mathbb{C}^n\rightarrow \mathbb{C}^m$
will be denoted $A_{i;j}$, so that $w=\hat{A}v$ is simply written
as $w_{i}=\hat{A}_{i;j}v_{j}$. Thus the `hat' operation acts as
follows:
\begin{equation}
\textrm{if} \;\; A\equiv A_{ij}\;\;\textrm{then}\;\;\hat{A}\equiv
\hat{A}_{i;j} \;\;\textrm{with} \;\; \hat{A}_{i;j}=A_{ij}
\end{equation}
Another useful interpretation of this operation is provided in
\cite{Verstraete}, by considering the canonical maximally
entangled state of $\mathbb{C}^n\otimes \mathbb{C}^n$,
$|\beta\rangle=\sum |j\rangle|j\rangle$. Indeed we have:
\begin{align}
A&=(\hat{A}\otimes Id_n)|\beta\rangle\label{verhat}\\
\hat{A}&=(Id_m\otimes\langle\beta|)(A\otimes Id_n)\nonumber
\end{align}
We now use the previous isomorphism to relate elements of
$M_{mn}(\mathbb{C})$ to linear maps from $M_n(\mathbb{C})$ to
$M_m(\mathbb{C})$. This formalizes some of the key steps in \cite{Su}\cite{Choi}\cite{Jamio}.
\begin{Iso}
\label{isomorphism2} The following linear map:
\begin{align}
\lonewh\;:\mathbb{C}^{mn}\otimes
(\mathbb{C}^{mn})^{\dagger} &\longrightarrow{End(M_{n}(\mathbb{C})\rightarrow M_{m}(\mathbb{C}))}\nonumber\\
\$&\longmapsto{[\whdoll:\rho\mapsto\whdoll(\rho)]}\nonumber\\
\textrm{such that} \;\; A B^{\dagger}
&\longmapsto{[\rho\mapsto\hat{A}\rho\hat{B}^{\dagger}]}\quad\textrm{i.e.}
\nonumber\\
\sum_{ijkl}
A_{ij}B^*_{kl}\ket{i}\ket{j}\bra{k}\bra{l}&\longmapsto{[\rho\mapsto
\sum_{ijkl}
A_{ij}B^*_{kl}\ket{i}\bra{j}\rho\ket{l}\bra{k}\;]}\nonumber
\end{align}
where $i,k=1, \ldots,m$ and $j,l=1,\ldots, n$, is an isomorphism. It is isometric in the sense that:
\begin{equation}
\label{isometry2} \forall \; \$, \europ \in M_{mn}(\mathbb{C}),
\;\;
\trace(\europ^{\dagger}\$)=\sum_{jl}\trace\big({\wheurop(E_{jl})}^{\dagger}\,\whdoll(E_{jl})\big),
\end{equation}
where $\{E_{jl}=\ket{j}\bra{l}\}$ is the canonical basis of
$M_n(\mathbb{C})$.
\end{Iso}
Before we give a proof we shall reassert Sudarshan's notation in
this case. Suppose $\$=\$_{ijkl}\ket{i}\ket{j}\bra{k}\bra{l}$ so
that we can write $\$\equiv \$_{ij;kl}$. We then have:
\begin{align}
& \whdoll\equiv \whdoll_{ik;jl} \; \;
\textrm{with}\;\;\whdoll_{ik;jl}=\$_{ij;kl}
\label{shulonghat}\\
&\textrm{so that}\;\;\whdoll:\rho_{j;l}\mapsto
\whdoll(\rho)_{i;k}=\whdoll_{ik;jl}\rho_{j;l}\nonumber
\end{align}
This notation views $End(M_{n}(\mathbb{C})\rightarrow
M_{m}(\mathbb{C}))$ as $m^2\times n^2$ matrices, or as
\emph{superoperators}, thus admitting the
usual Hilbert-Schmidt inner-product:
\begin{equation}
\trace(({\wheurop}^{\dagger}_{jl;ik})(\whdoll_{i'k';j'l'}))
\label{shuisometry2}
\end{equation}
where ${\wheurop}^{\dagger}_{jl;ik}$ is an $n^2 \times m^2$
matrix. The superoperator formalism simply consists of labelling a linear
operator on matrices by a super-matrix, or more generally a linear
map on tensors by a bigger tensor, and hence helps define operator norms.
 In fact  it will turn out to be a
corner stone of the state-operator correspondance. It has had many
applications in physics, amongst them the super-scattering or
``dollar'' operator 
formalism 
introduced in Quantum Field Theory by Hawking \cite{Hawking}, which,
in contrast with the S-matrix formalism, allows non-unitary evolutions (hence
our notation). \\
\emph{Proof of Isomorphism \ref{isomorphism2}.} Elements of
$\mathbb{C}^{mn}\otimes (\mathbb{C}^{mn})^{\dagger}$ are all of
the form $\sum_x A_x B_x^{\dagger}$, and thus by linearity the map
$\lonewh$ is fully determined by the above. The fact that it is an
isomorphism is made obvious by Equation
(\ref{shulonghat}). \\
Now let $\europ\equiv\europ_{ij;kl}$ and ${\$}={\$}_{ij;kl}$. We
now show that the notion of inner product given by
(\ref{shuisometry2}) is precisely that of the RHS of Equation
(\ref{isometry2}). Since
$\wheurop_{ik;jl}={\wheurop(E_{jl})}_{i;k}$ and
${\wheurop}^{\dagger}_{jl;ik}={\wheurop}^{*}_{ik;jl}$, we have
\begin{align*}
\trace(({\wheurop}^{\dagger}_{jl;ik})(\whdoll_{i'k';j';l'}))
&={\wheurop}^{\dagger}_{jl;ik}\whdoll_{ik;jl}\\
&=\sum_{ikjl}{\wheurop(E_{jl})}^{\dagger}_{k;i}{\whdoll(E_{jl})}_{i;k}\\
&=\sum_{jl}\trace({\wheurop(E_{jl})}^{\dagger}\,\whdoll(E_{jl}))\\
\end{align*}
Finally notice that
${\wheurop}^{\dagger}_{jl;ik}={\wheurop}^{*}_{ik;jl}=\europ^{*}_{ij;kl}$,
using (\ref{shulonghat}). Thus (\ref{shuisometry2}) is also equal
to the LHS of (\ref{isometry2}):
\begin{align*}
\trace(({\wheurop}^{\dagger}_{jl;ik})(\whdoll_{i'k';j';l'}))
&={\wheurop}^{\dagger}_{jl;ik}\whdoll_{ik;jl}\\
&=\europ^*_{ij;kl}\$_{ij;kl}=\europ^{\dagger}_{kl;ij}\$_{ij;kl}\\
&=\trace(\europ^{\dagger}\$) \quad \quad \Box
\end{align*}

In terms of the canonical maximally entangled state $\ket{\beta}$ of
$\mathbb{C}^n\otimes\mathbb{C}^n$, using (\ref{verhat}), we have that
\begin{equation}
\$=(\whdoll\otimes
Id_n)(|\beta\rangle\langle\beta|)\label{verlonghat}
\end{equation}
Note that $|\beta\rangle\langle\beta|=\sum E_{jl}\otimes E_{jl}$,
so we get
\begin{equation}
\label{dollarmatrix} \$=\sum_{jl}\whdoll(E_{jl})\otimes E_{jl}
\end{equation}
This relation is quite handy when one seeks to visualize the
isomorphism in terms of matrix manipulation. It is clear that the
isomorphisms $\hat{\phantom{U}}$ and  $\lonewh$ are biassed
towards interpreting states in $\mathbb{C}^{mn}=\mathbb{C}^m
\otimes \mathbb{C}^n$ as linear operations from states in the
second subspace $\mathbb{C}^n$ into states in the first subspace
$\mathbb{C}^m$. This will be made explicit in the forthcoming
theorems. Without difficulty we could do the contrary and view
states in $\mathbb{C}^{mn}$ as operations from
states in $\mathbb{C}^m$ to states in $\mathbb{C}^n$:\\
For $A=\sum_{ij}A_{ij}\ket{i}\ket{j}\in \mathbb{C}^{mn}$, let $\check{A}=\sum_{ij}
A_{ij}\ket{j}\bra{i}$, i.e. $\check{A}\equiv \check{A}_{j;i}=A_{ij}$,
so that $\check{A}=\hat{A}^t$. For
$\$=AB^{\dagger}\in M_{mn}(\mathbb{C})$ let
 $\wcdoll:M_m(\mathbb{C})\to M_n(\mathbb{C})$, $\rho\mapsto \sum_{ijkl}
A_{ij}B^*_{kl}\ket{j}\bra{i}\rho\ket{k}\bra{l}$, which implies:
\begin{equation}
\label{shutranspose} \wcdoll \equiv
\wcdoll_{jl;ik}=\$_{ij;kl}=\whdoll_{ik;jl}, \;\;\textrm{i.e.} \;\;
\wcdoll_{jl;ik}= \whdoll^{^t}_{jl;ik}.
\end{equation}
In this case Equation (\ref{dollarmatrix}) becomes:
\begin{equation}
\$=\sum_{ik}E_{ik}\otimes
\wcdoll(E_{ik})\label{reversedollarmatrix}
\end{equation}
Note that with the usual tensor product convention of taking the
right-hand-side matrix as the one to be plugged into each
component of the left-hand-side matrix, Equation
(\ref{reversedollarmatrix}) is simply written
$\$=(\wcdoll(E_{ik}))_{ik}$, which is precisely Choi's formalism in \cite{Choi}.
Thus many view these two Isomorphisms as rearrangements of the
coordinates of vectors or matrices. Although all would work
equally well with $\lonewc$, from now on we shall keep to our
initial version of the isomorphisms, taking the second subspace
into the first.
\\

\subsection{Useful Formulae}

The following two lemmas are simple but useful results related to
isomorphisms $1$ and $2$.

\begin{Lem}
Let $A,B \in \mathbb{C}^m\otimes \mathbb{C}^n$, so that
$AB^{\dagger} \in \mathbb{C}^{mn}\otimes
(\mathbb{C}^{mn})^{\dagger}$, and let $\trace_1$ and $\trace_2$
denote the partial traces on $\mathbb{C}^m$ and $\mathbb{C}^n$
respectively. Then we have:
\begin{align}
\trace_1(AB^{\dagger})&=({\hat{B}}^{\dagger}\hat{A})^t  \label{trace1}\\
\trace_2(AB^{\dagger})&=\hat{A}{\hat{B}}^{\dagger} \label{trace2}
\end{align}
\end{Lem}
\emph{Proof.} let $A \equiv A_{ij}$ and $B \equiv B_{kl}$ with
$i,k=1,\ldots, m$ and $j,l=1,\ldots, n$.
$AB^{\dagger}=A_{ij}B_{kl}^*\ket{i}\bra{k}\otimes \ket{j}\bra{l}$.
Thus taking $\trace_1$ sets $i=k$ and taking $\trace_2$ sets
$j=l$:
\begin{align*}
\trace_1(AB^{\dagger})_{j;l}=A_{ij}B^*_{il}=
{\hat{B}}^{\dagger}_{l;i}\hat{A}_{i;j}={(\hat{B}^{\dagger}\hat{A})^t}_{j;l}
\\
\trace_2(AB^{\dagger})_{i;k}=\hat{A}_{ij}B_{kj}^*={\hat{A}{\hat{B}}^{\dagger}}_{i;k}
\quad \Box
\end{align*}

\begin{Lem}\label{trace hat}
Suppose $\lonewh$ is defined for $n$ fixed and for all $d$ such
that it takes any element of $\mathbb{C}^{dn}\otimes
(\mathbb{C}^{dn})^{\dagger}$ to a linear map from
$M_n(\mathbb{C})$ to $M_d(\mathbb{C})$:
\begin{equation*}
\forall d, \quad \lonewh :\mathbb{C}^{dn}\otimes
(\mathbb{C}^{dn})^{\dagger} \longrightarrow
End(M_{n}(\mathbb{C})\rightarrow M_{d}(\mathbb{C})),
\end{equation*}
and let $\trace_1$ denote the partial trace on the first
$r$-dimensional subsystem of any system. We then have:
\begin{equation*}
\forall \;\$\; \in \mathbb{C}^{rmn} \otimes
(\mathbb{C}^{rmn})^{\dagger},\quad \whx{\trace_1(\$)}=\trace_1
\circ \whdoll
\end{equation*}
in other words $\trace_1$ and $\lonewh$ commute.

\end{Lem}
\emph{Proof:} In the following, $i,k= 1,\ldots, m $ and  $j,l=1,
\ldots, n $ as usual, while $p,q =1, \ldots, r$. Let $\$\equiv
\$_{pij;qkl} \in \mathbb{C}^{rmn} \otimes (\mathbb{C}^{rmn})^*$,
and $\rho = \rho_{j;l}\in
M_n(\mathbb{C})$.\\
Then
$\whdoll(\rho)_{pi;qk}=\whdoll_{piqk;jl}\rho_{j;l}=\$_{pij;qkl}\rho_{j;l}$
is in $M_{rm}(\mathbb{C})$. Since $\trace_1$ sets $p=q$,
${(\trace_1 \circ \whdoll)(\rho)}_{i;k}= \$_{pij;pkl}\rho_{j;l}$.
On the other hand ${\trace_1(\$)}_{ij;kl}= \$_{pij;pkl}$ so
${\whx{\trace_1(\$)}}\equiv
{\whx{\trace_1(\$)}}_{ik;jl}=\$_{pij;pkl}$, thus
$\whx{\trace_1(\$)}(\rho)_{i;k}=\$_{pij;pkl}\rho_{j;l}$.
$\quad\Box$ \\

Next we give a novel and powerful formula relating linear
operations $\whdoll$ to trace outs of matrix multiplications
involving $\$$.

\begin{Proposition} \label{multiplication}
Let $\whdoll$ a linear map from $M_n(\mathbb{C})$ to
$M_m(\mathbb{C})$, $\sigma$, $\rho$ two elements of
$M_n(\mathbb{C})$, $\kappa$, $\tau$ two elements of
$M_m(\mathbb{C})$. Then we have:
\begin{align}
\label{bigmultiplication} \kappa\whdoll(\rho \sigma)\tau&=
\trace_2\big( (\kappa\otimes
\rho^t)\$(\tau\otimes \sigma^t)\big)
\end{align}
where $\trace_2$ denotes the partial trace over the second system
$\mathbb{C}^n$  in
$\mathbb{C}^m \otimes \mathbb{C}^n$. This implies that for all $\rho
\in M_n(\mathbb{C})$ and $\kappa \in M_m(\mathbb{C})$, 
\begin{equation}
\trace\big(\kappa\whdoll(\rho)\big)= \trace\big((\kappa\otimes
\rho^t)\$\big). \label{totaltrace}
\end{equation}
\end{Proposition}
\emph{Proof:} Since $(\kappa\otimes
\rho^t)_{ij;kl}=\kappa_{ik}\rho^t_{jl}$, $(\tau\otimes
\sigma^t)_{ij;kl}=\tau_{ik}\sigma^t_{jl}$, and tracing out
$\mathbb{C}^n$ consists of setting $j=l$, we have
\begin{align*}
(\kappa\otimes \rho^t)\$(\tau\otimes
\sigma^t)_{ij;kl}&=\kappa_{ii'}\rho^t_{jj'}\$_{i'j';i''j''}\tau_{i''k}\sigma^t_{j''l}
\\
\trace_2\big( (\kappa\otimes \rho^t)\$(\tau\otimes
\sigma^t)\big)_{i;k}&=\kappa_{ii'}\rho_{j'l}\$_{i'j';i''j''}\tau_{i''k}\sigma_{lj''}\\
 &=\kappa_{ii'}\$_{i'j';i''j''}\rho_{j'l}\sigma_{lj''}\tau_{i''k} \\
  &=\kappa_{ii'}\whdoll_{i'i'';j'j''}(\rho_{j'l}\sigma_{lj''})\tau_{i''k}\\
  &=\kappa\whdoll(\rho \sigma )\tau.
\end{align*}
Equation (\ref{totaltrace}) follows immediately by letting
$\tau=Id_m$, $\sigma=Id_n$ and taking the total trace.
 $\Box$ \\
From Equation (\ref{bigmultiplication})  one can also derive the
following interesting formula:  $\forall \rho \in
M_n(\mathbb{C})$,
\begin{align}
&\whdoll\big((\rho^{\dagger}\rho)^t\big) =
\trace_2\big((Id_m\otimes\rho)\$(Id_m\otimes
\rho^{\dagger})\big)\label{dollarmment}
\end{align}
We shall come back to Equation (\ref{dollarmment}) in subsection
\ref{duality}, with a more physical point of view. For now note
that the equation is slightly more general than the one given in
\cite{Verstraete}p4, and that its equivalent form for $\lonewc$ is
clearly seen to define a map from the first subspace into the
second:
\begin{equation*}
\wcdoll\big((\rho^{\dagger}\rho)^t\big)=\trace_1\big((\rho\otimes
Id_n)\$(\rho^{\dagger}\otimes Id_n)\big).
\end{equation*}
Moreover the original Equation (\ref{totaltrace}) will
have a wide range of applications in the field of quantum
information theory. This is because many of the mathematical
problems raised by quantum cryptography require a careful
optimization of the fidelities induced by a linear operator
$\whdoll$. By means of this formula such involved expressions can
elegantly be brought to just the trace of the product of two
matrices \cite{decoys}. 

\subsection{The correspondence}
\label{the correspondence} We proceed to give the well-known three
fundamental theorems about isomorphism $2$.  
\begin{Th}
\label{herm-preserving}
The linear operation $\whdoll : M_n(\mathbb{C}) \rightarrow
M_m(\mathbb{C})$ is \emph{Hermitian-preserving} if and only if
$\$$ belongs to $\textrm{Herm}_{mn}(\mathbb{C})$.
\end{Th}
\emph{Proof.} $[\Rightarrow]$ Suppose $\whdoll$
Hermitian-preserving, then by Remark \ref{Hermextends} so is
$(\whdoll\otimes Id_n)$. Now since $|\beta\rangle\langle\beta|$ is
hermitian it must be the case that $(\whdoll\otimes
Id_n)(|\beta\rangle\langle\beta|)=\$$ is hermitian. We used
Equation (\ref{verlonghat}) for the last
equality.\\
$[\Leftarrow]$ Suppose $\$$ Hermitian, so that
$\$_{ij;kl}=\$_{kl;ij}^*$. Let $\rho_{jl}=\rho_{lj}^* \in
\textrm{Herm}_n(\mathbb{C})$. Using (\ref{shulonghat}) we have
\begin{align*}
\whdoll(\rho)_{i;k}&=\whdoll_{ik;jl}\rho_{jl}=\$_{ij;kl}\rho_{jl} \\
 &=\$^*_{kl;ij}\rho^*_{lj}=(\whdoll_{ki;lj}\rho_{lj})^* \\
 &= \whdoll(\rho)^*_{k;i}
\end{align*}
so that $\whdoll$ is Hermitian-preserving. $\Box$ \\
This result first appeared in \cite{depillis}. In terms of
components, $\whdoll$ is Hermitian-preserving if and
only if $\$_{ij;kl}=\$_{kl;ij}^*$, or equivalently
$\whdoll_{ik;jl}={\whdoll_{ki;lj}}^*$. \\

\begin{Th}
\label{Positive-preserving} The linear operation $\whdoll :
M_n(\mathbb{C}) \rightarrow M_m(\mathbb{C})$ is
\emph{Positive-preserving} if and only if $\$$ belongs to
$\textrm{Herm}_{mn}(\mathbb{C})$ and is such that for all
separable state $\rho$ in $\textrm{Herm}_{mn}^+(\mathbb{C})$,
$\trace(\$ \rho)\geq 0$.
\end{Th}
\emph{Proof:} $\$$ is Hermitian by theorem \ref{herm-preserving} since $\whdoll$ is
Hermitian-preserving. Using Equation (\ref{totaltrace}) in the
following, with $\rho, \rho_1\in \textrm{Herm}^+_n(\mathbb{C})$
and $\sigma, \rho_2\in \textrm{Herm}^+_m(\mathbb{C})$, we have:
\begin{align*}
&\whdoll \;\; \textrm{is Positive-preserving} \\
\Leftrightarrow & \forall \rho, \;\;\forall \sigma, \;\;
\trace\big(\sigma\whdoll(\rho)\big) \geq 0 \\
\Leftrightarrow & \forall \rho, \;\;\forall \sigma, \;\;
\trace\big((\sigma\otimes \rho^t)\$\big) \geq 0 \\
\Leftrightarrow & \forall\rho_1, \;\; \forall \rho_2, \;\;
\trace\big(\$ (\rho_1\otimes \rho_2)\big) \geq 0 \\
\Leftrightarrow & \forall \rho \in
\textrm{Herm}_{mn}^+(\mathbb{C})\;\;\textrm{separable}, \;\; \trace(\$
\rho)\geq 0  \quad \Box
\end{align*}
This result is shown for instance in \cite{Horodeckis}, in a
different manner. We shall come back to its geometrical consequences
in section \ref{geometrical structure}. \\


\begin{Th}
\label{operations as states} The linear operation $\whdoll :
M_n(\mathbb{C}) \rightarrow M_m(\mathbb{C})$ is \emph{Completely
Positive-preserving} if and only if $\$$ belongs to
$\textrm{Herm}_{mn}^+(\mathbb{C})$.
\end{Th}
\emph{Proof.} $[\Rightarrow]$ Suppose $\whdoll$ Completely
Positive-preserving. Since $|\beta\rangle\langle\beta|$ is
positive it must be the case that $(\whdoll\otimes
Id_n)(|\beta\rangle\langle\beta|)=\$$ is positive. We used
Equation (\ref{verlonghat}) for the last
equality.\\
$[\Leftarrow]$ Suppose $\$$ positive. 
We want to show that for all $r$, $\whdoll \otimes Id_r:
M_{nr}(\mathbb{C}) \to M_{mr}(\mathbb{C})$ is
Positive-preserving. Let $\europ \in M_{(mr)(nr)}(\mathbb{C})$
be such that:
\begin{equation*}
\wheurop = {\whdoll \otimes Id_r}.  
\end{equation*}
Explicitely, with $s,t,u,v=1, \ldots ,r$, and $i,k=1, \ldots,m$ and $j,l=1,
\ldots ,n$ as usual, 
\begin{align}
(\whdoll \otimes
  Id_r)_{(is)(kt);(ju)(lv)}& =\delta_{su}\delta_{tv}\whdoll_{ik;jl}
  \nonumber \\
&=\delta_{su}\delta_{tv}\$_{ij;kl} \label{eurocomp}  \\          
&= \wheurop_{(is)(kt);(ju)(lv)}  \nonumber \\
           &= \europ_{(is)(ju);(kt)(lv)} \nonumber
\end{align}
where we have used (\ref{shulonghat}) to switch from $\whdoll$ to $\$$
and $\wheurop$ to $\europ$.
Let $V_{(kt)(lv)} \in \mathbb{C}^{(mr)(nr)}$. Using (\ref{eurocomp})
and the fact that $\$ \in \textrm{Herm}_{mn}^+(\mathbb{C})$, we
get   
\begin{align}
V^{\dagger }\europ V
 &=V^*_{(is)(ju)}\europ_{(is)(ju);(kt)(lv)}V_{(kt)(lv)}  \nonumber \\
 &=V^*_{isjs}\$_{ij;kl}V_{ktlt} \nonumber \\
&\geq 0 \nonumber, 
\end{align}
hence $\europ \in
 \textrm{Herm}_{(mr)(nr)}^+(\mathbb{C})$.    
Then, by theorem \ref{Positive-preserving},  $\whdoll \otimes Id_r$
is Positive-preserving if for all $\rho_1 \in
\textrm{Herm}_{rm}^+(\mathbb{C})$ and $\rho_2 \in
\textrm{Herm}_{rn}^+(\mathbb{C})$, 
$\trace(\europ (\rho_1\otimes \rho_2))\geq 0$. This follows directly
 since $\rho_1 \otimes \rho_2$ and $\europ$ are positive. $\Box$ \\
This result first appears in \cite{Su}, and later \cite{Choi} with a different proof. The (possibly
non-normalized) states of a $mn$-dimensional quantum system, or
elements of $\textrm{Herm}^+_{mn}(\mathbb{C})$, are thus in
one-to-one correspondence with the (possibly non Trace-preserving)
quantum operations, or Completely Positive-preserving maps, taking
an $n$-dimensional system into an $m$-dimensional system. We claim
that virtually all of the important, well-established results
about quantum operations are in direct correspondence with those
regarding quantum states, through the use of theorem
\ref{operations as states}. In \cite{Su}\cite{Choi}\cite{Landau}\cite{Shu},
 the operator 
sum representation for Completely Positive-preserving maps is derived in the
 proof of theorem \ref{operations as states}, but in our approach we
 will think of it as stemming directly from the properties of quantum
states.

\section{Properties of quantum states and quantum operations}

\subsection{Properties rediscovered via the correspondence}
\label{properties}

\begin{Pro}[Decomposition, degree of freedom.] \label{state decomposition}
A matrix $\rho$ is in $\textrm{Herm}_{d}^+(\mathbb{C})$ if and
only if it can be written as
\begin{equation*}
\rho=\sum_x A_x A^{\dagger}_x
\end{equation*} where each $A_x$ is a
$d$-dimensional vector. Two decompositions $\{A_x\}$ and $\{B_y\}$
correspond to the same state $\rho$ if and only if there exists an
isometric matrix $U$ (i.e. $U^{\dagger}U=Id$) such that $A_x=\sum
U_{xy}B_y$. There is a decomposition $\{A_x\}$ with
$rank(\rho)\leq d$ non-zero elements and such that
$A^{\dagger}_{x'} A_{x} \propto \delta_{xx'}$.
\end{Pro}
\begin{Cor}[Operator sum representation.]
\label{operator decomposition} A linear map
$\whdoll:M_{n}(\mathbb{C})\rightarrow M_{m}(\mathbb{C})$ is
Completely Positive-preserving if and only if it can be written as
\begin{equation*}\whdoll:\;\rho\mapsto
\sum_x \hat{A}_x \rho \hat{A}^{\dagger}_x
\end{equation*}
where each $\hat{A}_x$ is an $m\times n$ matrix. Two
decompositions $\{\hat{A}_x\}$ and $\{\hat{B}_y\}$ correspond to
the same $\whdoll$ if and only if there exists an isometric matrix
$U$ (i.e. $U^{\dagger}U=Id$) such that $\hat{A}_x=\sum
U_{xy}\hat{B}_y$. There is a decomposition $\{\hat{A}_x\}$ with
$r\leq mn$ elements and such that $\trace(\hat{A}^{\dagger}_{x'}
\hat{A}_x) \propto \delta_{xx'}$. $r$ will be referred to as the
higher rank rank of $\whdoll$, as this is the decomposition having the
least number of elements.
\end{Cor}
\emph{Proof of Property \ref{state decomposition}.} This is the
spectral decomposition theorem for positive matrices, together with
the unitary degree of freedom theorem by Hughston, Josza and Wooters
\cite{Nielsen}p103. $\quad\Box$\\
\emph{Proof of Corollary \ref{operator decomposition}.} Consider
$\whdoll$ a Completely Positive-preserving linear operator. By
theorem \ref{operations as states}, $\$$ is positive, and so
Property \ref{state decomposition} provides decompositions upon
that state. One may translate back these decompositions in terms
of quantum operations using Isomorphism $2$: this yields nothing
but Corollary
\ref{operator decomposition}. $\quad\Box$\\
Notice that the higher rank of $\whdoll$ is equal to $rank(\$)$.
\begin{Pro}[Purification.] \label{state purification}
A matrix $\rho$ is in
$\textrm{Herm}_{d}^+(\mathbb{C})$ if and only if it can be written
as
\begin{equation*}
\rho=\trace_1(\rho_{pure})\quad\textrm{with}\quad\rho_{pure}=V
V^{\dagger}
\end{equation*}
where $V$ is an $rd$-dimensional vector and $\trace_1$ traces out
the first  $r$-dimensional subsystem ($r$ can be chosen equal to
$rank(\rho)\leq d$).
\end{Pro}
\begin{Cor}[Factorizable then trace representation.] \label{operator purification}
A linear map $\whdoll:M_{n}(\mathbb{C})\rightarrow
M_{m}(\mathbb{C})$ is Completely Positive-preserving if and only
if it can be written as
\begin{equation*}
\whdoll:\rho\mapsto \trace_1(\whdoll_{pure}(\rho))
\quad\textrm{with}\quad\whdoll_{pure}:\rho\mapsto \hat{V}\rho
\hat{V}^{\dagger}
\end{equation*}
where $\hat{V}$ is an $rm \times n$ matrix and $\trace_1$ traces
out the first $r$-dimensional subsystem ($r$ may be chosen equal
to $rank(\whdoll)\leq mn$). Moreover if $\,\whdoll$ decomposes as
$\{\hat{A}_x\}$ we have:
\begin{equation} \label{purification and trace}
\hat{V}^{\dagger}\hat{V}=\sum_x \hat{A}^{\dagger}_x\hat{A}_x
\end{equation}
\end{Cor}
\emph{Proof of Property \ref{state purification}.} $[\Rightarrow]$
Suppose $\rho$ decomposes as $\{A_x\}$ and let $V=\sum
\ket{x}A_x$, with $\{\ket{x}\}$ an orthonormal basis of an ancilla system.
\begin{align*}
\rho_{pure}&=VV^{\dagger}\\
&=\sum_{xy} \ket{x}\bra{y}\otimes A_x A^{\dagger}_y\\
\trace_1(\rho_{pure})&=\sum_{xy}\langle y\ket{x} A_x A^{\dagger}_y\\
&=\sum_x A_x A^{\dagger}_x=\rho
\end{align*}
If $\{A_x\}$ is a spectral decomposition of $\rho$ it counts $rank(\rho)$
elements, and thus $r$ can be chosen to equal $rank(\rho)$.
\begin{align*}
[\Leftarrow]\quad\quad\quad \bra{\psi}\rho\ket{\psi}&=\sum_i
\bra{i}\bra{\psi}VV^{\dagger}\ket{i}\ket{\psi}\geq 0
\quad\textrm{since}\\
\forall i \quad
&\bra{i}\bra{\psi}VV^{\dagger}\ket{i}\ket{\psi}\geq 0
\quad\quad\quad\Box
\end{align*}
The second corollary is not traditionally thought of as a `quantum
operation equivalent' of quantum state purification. We now
explicitly show how the result is again trivially obtained from
Property \ref{state purification}, by virtue of Theorem
\ref{operations as states}.\\
\emph{Proof of Corollary \ref{operator purification}.} Consider
$\whdoll$ a Completely Positive-preserving linear operator. By
Theorem \ref{operations as states}, $\$$ is positive, and so
Property \ref{state purification} gives $\$=\trace_1(\$_{pure})$,
$\$_{pure}=VV^{\dagger}$, where the ancilla system can be chosen
to be of dimension $r=rank(\$)$. As a consequence we can use Lemma
\ref{trace hat} to retrieve $\whdoll=\trace_1(\whdoll_{pure})$,
$\whdoll_{pure}:\rho\mapsto\hat{V}\rho\hat{V}^{\dagger}$.\\
Moreover, denote by  $\trace_{1'}$ the partial trace over the
$m$-dimensional system. For $V=V_{xij}\ket{x}\ket{i}\ket{j}$, let
$\hat{V}\equiv V_{xij}\ket{x}\ket{i}\bra{j}$ the corresponding $rm
\times n$ matrix. Since $\trace_1(\rho_{pure})=\rho$ with
$\rho_{pure}=V V^{\dagger}$, and $\rho=\sum_x A_x A_x^{\dagger}$,
we get
\begin{align*}
(\trace_{1'}\circ \trace_1)(V V^{\dagger}) &=\sum_x\trace_{1'}(A_x
A_x^{\dagger})
\quad\textrm{implying}\\
\hat{V}^{\dagger}\hat{V} &=\sum_x \hat{A}_x^{\dagger} \hat{A}_x
\quad\textrm{by Equation (\ref{trace1})} \;\;\Box
\end{align*}
Notice that whenever $\whdoll$ is Trace-preserving,  then Equation
(\ref{purification and trace}) reads
$\hat{V}^{\dagger}\hat{V}=Id_n$, so that $\hat{V}$ is isometric.
Thus we have derived as a simple consequence of properties of
state purification that any Trace-preserving quantum operation can
arise as the trace-out of an isometric operation.

\begin{Pro}[Schmidt decomposition.]
\label{state schmidt} Consider $\rho=VV^{\dagger}$ a
non-normalized pure state in $\textrm{Herm}_{mn}^+(\mathbb{C})$
with $V=\sum V_{ij }\ket{i}\ket{j}$ in the canonical basis of
$\mathbb{C}^m\otimes \mathbb{C}^n$. Then there exists some
positive reals $\{\lambda_i\}$ and some orthogonal basis
$\{\ket{\psi_i}\}$ and $\{\ket{\phi_i}\}$ of $\mathbb{C}^m$ and
$\mathbb{C}^n$ respectively, such that
\begin{equation*}
V=\sum_{i=1}^r \lambda_i \ket{\psi_i} \ket{\phi_i},
\end{equation*}
with $r\leq m$ and $r\leq n$. Moreover:
\begin{align*}
&\trace_1(\rho)=\sum_{i=1}^r \lambda_i^2 \ket{\phi_i}
\bra{\phi_i}\quad\;(n\times n\;\textrm{positive})\\
&\trace_2(\rho)=\sum_{i=1}^r \lambda_i^2 \ket{\psi_i}
\bra{\psi_i}\quad(m\times m\;\textrm{positive})
\end{align*}
\end{Pro}
\begin{Cor}[Polar Decomposition.] \label{operator schmidt}
Consider $\whdoll:M_n(\mathbb{C})\rightarrow M_m(\mathbb{C}),
\,\rho\mapsto\hat{V}\rho\hat{V}^{\dagger}$ a factorizable
Completely Positive-preserving linear map, with $\hat{V}=\sum
V_{ij}\ket{i}\bra{i}$. Then there exists some positive reals
$\{\lambda_i\}$ and some orthogonal basis of $\mathbb{C}^m$ and
$(\mathbb{C}^n)^{\dagger}$ , namely $\{\ket{\psi_i}\}$ and
$\{\bra{\phi_i^*}\}$, such that
\begin{equation*}
\hat{V}=\sum_{i=1}^r \lambda_i \ket{\psi_i} \bra{\phi_i^*}
\end{equation*}
with $r\leq m$ and $r\leq n$. In other words:
\begin{align*}
\hat{V}&=UJ=KU \quad\textrm{with}\\
J&=\sqrt{\hat{V}^{\dagger}\hat{V}}=\sum_{i=1}^r \lambda_i \ket{\phi_i^*} \bra{\phi_i^*}\quad(n\times n\;\textrm{positive})\\
K&=\sqrt{\hat{V}\hat{V}^{\dagger}}=\sum_{i=1}^r \lambda_i \ket{\psi_i} \bra{\psi_i}\quad(m\times m\;\textrm{positive})\\
U&=\sum_{i=1}^n\ket{\psi_i}\bra{\phi_i^*}\quad\;(m\times
n\;\textrm{isometric, i.e.}\; U^{\dagger}U=Id_n)
\end{align*}
\end{Cor}
\emph{Proof of Property \ref{state schmidt}.}
Let $\rho=VV^{\dagger}$, $V=\sum V_{ij}\ket{i}\ket{j}$, and $\trace_2$
the partial trace on the last $n$-dimensional system. Since
$\rho^A=\trace_2(\rho)$ is in $\textrm{Herm}^+_{m}(\mathbb{C})$, we can
write
\begin{equation*}
\rho^A=\sum_{i=1}^r \lambda_i^2 \ket{\psi_i}\bra{\psi_i}
\end{equation*}
where $\{\lambda_i\}$ are strictly positive reals, $r\leq m$, and
$\{\ket{\psi_i}\}$ is an orthonormal family of vectors which we
may complete into an orthonormal basis of $\mathbb{C}^m$. By
expressing the first subspace of $V$ in this basis we can of
course write:
\begin{equation*}
V=\sum_{i=1}^r
\ket{\psi_i}\ket{\tilde{\phi}_i}\quad\textrm{with}\quad\ket{\tilde{\phi}_i}=(\bra{\psi_i}\otimes
Id_n)V
\end{equation*}
We have:
\begin{align*}
\bra{\tilde{\phi}_i}\tilde{\phi}_j\rangle
&=\trace(\ket{\tilde{\phi}_j}\bra{\tilde{\phi}_i})\\
&=\trace((\bra{\psi_i}\otimes Id)VV^{\dagger}(\ket{\psi_j}\otimes
Id))\\
&=\trace((\ket{\psi_j}\bra{\psi_i}\otimes Id)VV^{\dagger})\\
&=\trace(\ket{\psi_j}\bra{\psi_i}\rho^A)\\
&=\lambda_i^2 \delta_{ij}
\end{align*}
Thus $\{\ket{\phi_i}=\ket{\tilde{\phi}_i}/\lambda_i \}$ is an
orthonormal family of vectors in $\mathbb{C}^n$, which we may again
complete into an orthonormal basis.\\
We now have $V=\sum \lambda_i \ket{\psi_i}\ket{\phi_i}$ , from which it is
straightforward to verify that
\begin{equation*}
\trace_1(\rho)=\sum_{i=1}^r \lambda_i^2
\ket{\phi_i}\bra{\phi_i}\quad\Box
\end{equation*}

The well-known connection between the Schmidt decomposition and the polar
decomposition (itself trivially equivalent to the singular
value decomposition) is now shown to arise naturally using the
state-operator correpondence.\\
\emph{Proof of Corollary  \ref{operator schmidt}.} Consider
$\whdoll:\rho\mapsto\hat{V}\rho\hat{V}^{\dagger}$. Using
Isomorphism $2$ the corresponding state in
$\textrm{Herm}_{mn}^+(\mathbb{C})$ is $\rho=VV^{\dagger}$.
Applying the Schmidt decomposition theorem yields
\begin{align*}
V&=\sum_{i=1}^r \lambda_i \ket{\psi_i} \ket{\phi_i}\quad\textrm{and thus}\\
\hat{V}&=\sum_{i=1}^r \lambda_i \ket{\psi_i} \bra{\phi_i^*}
\end{align*}
with $\{\ket{\psi_i}\}$ and $\{\bra{\phi_i^*}=\bra{\phi_i}^*\}$
some orthogonal basis of $\mathbb{C}^m$ and
$(\mathbb{C}^n)^{\dagger}$ respectively. Now if we call $U$ the
$m\times n$ isometric (i.e. $U^{\dagger}U=Id_n$) matrix
$\sum_{i=1}^{n}\ket{\psi_i} \bra{\phi_i^*}$,  we have that
$\hat{V}=UJ=KU$, with
\begin{align*}
K&=\sum_{i=1}^r\lambda_i\ket{\psi_i}\bra{\psi_i}=\sqrt{\trace_2(VV^{\dagger})}=\sqrt{\hat{V}\hat{V}^{\dagger}}\\
J&=\sum_{i=1}^r
\lambda_i\ket{\phi_i^*}\bra{\phi_i^*}=\left(\sqrt{\trace_1(VV^{\dagger})}\right)^t=\sqrt{\hat{V}^{\dagger}\hat{V}}.
\end{align*}
In the above $K$ is $m\times m$ whilst $J$ is $n\times n$, and the
last equality of each line was derived from Equations
(\ref{trace2}) and (\ref{trace1}). $\quad\Box$\\

Thus it seems that all the standard results about quantum
operations are in correspondence with those concerning quantum
states. Of course although we derived the properties of operators from
those of states, we could equally have done the opposite. Next we seek
to apply the same principle to derive new
results, as we consider properties of states and operations which
do not yet have any equivalent in terms of, respectively,
operations and states.

\subsection{Properties discovered via the correspondence}
\label{new properties} We first derive a factorizability condition
on quantum operations by making use of the well-known property:
\begin{Pro}[Purity condition.] \label{state purity}
Let $\rho$ a matrix in $\textrm{Herm}_{d}^+(\mathbb{C})$. Then
$\rho$ is non-normalized pure, i.e. of the form
$\rho=VV^{\dagger}$, if and only if
\begin{equation*}
\trace(\rho)^2-\trace(\rho^2)=0
\end{equation*}
\end{Pro}
\begin{Cor}[Factorizability condition.] \label{operator purity}
Let $\whdoll:M_n(\mathbb{C})\rightarrow M_m(\mathbb{C})$ a
Completely Positive-preserving linear operator. Then $\whdoll$ is
of the form $\whdoll: \rho\mapsto \hat{V}\rho\hat{V}^{\dagger}$,
i.e. it is factorizable,  if and only if
\begin{equation}
\label{factorizability condition}
\left(\trace(\whdoll(Id_n))\right)^2
-\sum_{jl}\trace\big({\whdoll(E_{jl})}^{\dagger}\whdoll(E_{jl})\big)=0
\end{equation}
or equivalently in terms of indices
\begin{equation*}
(\whdoll_{ii;jj})^2 - \whdoll^*_{ik;jl}\whdoll_{ik;jl}=0.
\end{equation*}
\end{Cor}
\emph{Proof of Property \ref{state purity}.}$[\Rightarrow]$ is obvious
since $\rho$ pure has only got one non-zero eigenvalue.\\
$[\Leftarrow]$ Suppose $\rho$ has eigenvalues $\{\lambda_i\}$.
The purity condition amounts to
\begin{equation*}
(\sum_i \lambda_i)^2=\sum_i \lambda_i^2\quad\textrm{implying}
\quad\sum_{i<j} \lambda_i \lambda_j =0.
\end{equation*}
For the last relation to hold, since the $\lambda_i$'s are
positive
there can be at most one value of i such that $\lambda_i\neq0$.$\quad\Box$\\
\emph{Proof of Corollary \ref{operator purity}.} $\whdoll$ is
factorizable is equivalent to $\$$ being pure, thus by Property
\ref{state purity} to
\begin{equation*}
\trace(Id_{mn} \$)^2-\trace(\$^2)=0.
\end{equation*}
Using $\$^{\dagger}=\$$ Equation (\ref{factorizability condition})
is a direct application of Equation (\ref{isometry2}) upon this
last equation, as can be seen from
\begin{align*}
Id_{mn}&=\sum_{kl} \ket{kl}\bra{kl}\\
\textrm{so that} \;\;
\whx{Id_{mn}}&:\rho\mapsto\sum_{kl}E_{kl}\rho
E^{\dagger}_{kl}\\
\textrm{and} \;\; \whx{Id_{mn}}&:E_{jl}\mapsto\delta_{jl}\,Id_m
\quad\quad\Box
\end{align*}
Next we give two new vector decompositions which stem from classical
results on matrix decomposition.
\begin{Pro}[One-sided triangular decomposition.] \label{state QR}
Let $\rho=VV^{\dagger}$ a non-normalized pure state in
$\textrm{Herm}_{mn}^+(\mathbb{C})$, with $V=\sum
V_{ij}\ket{i}\ket{j}$ in the canonical basis, and suppose $m\geq
n$. Then there exists some orthogonal basis of $\mathbb{C}^m$,
namely $\{\ket{\psi_i}\}$, such that
\begin{equation*}
V=\sum_{i\leq j}^{j=n} \mu_{ij} \ket{\psi_i} \ket{j}
\end{equation*}
\end{Pro}
\emph{Proof.} According to the $QR$ decomposition theorem
\cite{matrices} the $m\times n$ matrix $\hat{V}$ can be decomposed
as $\hat{V}=QR$, where $Q$ is $m\times n$ and verifies
$Q^{\dagger}Q=Id_n$ whilst $R$ is $n \times n$ upper triangular.
Thus we have:
\begin{align*}
\hat{V}&=Q \sum_{i\leq j}^{j=n} \mu_{ij} \ket{i} \bra{j}\\
\hat{V}&= \sum_{i\leq j}^{j=n} \mu_{ij} \ket{\psi_i} \bra{j}\\
V&=\sum_{i\leq j}^{j=n} \mu_{ij} \ket{\psi_i} \ket{j}
\end{align*}
Since $Q$ is isometric,  the $\{\ket{\psi_i}=Q\ket{i} \}$ are othornomal
and can be extended to form a basis of $\mathbb{C}^m$.$\quad\Box$

On the one hand Propety \ref{state QR} is less powerful than the
Schmidt decomposition, in the sense that it yields `upper
triangular' coefficients $\mu_{ij}$ instead of the neat diagonal
form $V=\sum_{i}^{r} \lambda_{i} \ket{\psi_i} \ket{\phi_i}$. On
the other hand however our Property requires a change of basis for
the first subsytem only. Such a disctinction is perfectly
analoguous to what separates the polar decomposition (or more
expressively its singular value decomposition corollary) from the
$QR$ decomposition when speaking about matrices. Just like the
$QR$ decomposition the one-sided state triangularization is easily computed. \\
Schur's triangularization theorem can also be given a quantum state
equivalent, as we now exlpain. This seems of a lesser interest
however, since the procedure involves two changes of basis, one for
each subsystem - a case which seems better covered by the Schmidt
decomposition (though here the two basis are simply related).

\begin{Pro}[Two-sided triangular decomposition.] \label{state schur}
Let $\rho=VV^{\dagger}$ a non-normalized pure state in
$\textrm{Herm}_{m^2}^+(\mathbb{C})$, with $V=\sum V_{ij
}\ket{i}\ket{j}$ in the canonical basis. Then there exists some
orthogonal basis of $\mathbb{C}^m$, namely $\{ \ket{\psi_i} \}$
such that
\begin{equation*}
V=\sum_{i\leq j}^{j=m} \mu_{ij} \ket{\psi_i} \ket{\psi^*_j}
\end{equation*}
where ${\,}^*$ denotes complex conjugation of the coordinates of a
vector in the canonical basis. Moreover the set $\{\mu_{ii}\}$ is
the set of the Schmidt coefficients $\{\lambda_i\}$ of $V$ (as
defined in Property \ref{state schmidt}).
\end{Pro}
\emph{Proof.} According to Schur's decomposition theorem
\cite{matrices} the matrix $\hat{V}$ can be decomposed as
$\hat{V}=UTU^{\dagger}$, where $U$ is unitary and  $T$ is upper
triangular and has the singular values of $\hat{V}$ in the
diagonal (i.e. precisely the $\lambda_i$'s of the polar
decomposition and of the Schmidt decomposition). And so we have:
\begin{align*}
\hat{V}&=U \sum_{i\leq j}^{j=m} \mu_{ij} \ket{i} \bra{j} U^{\dagger}\\
\hat{V}&= \sum_{i\leq j}^{j=m} \mu_{ij} \ket{\psi_i} \bra{\psi_j}\\
V&= \sum_{i\leq j}^{j=m} \mu_{ij} \ket{\psi_i} \ket{\psi^*_j}
\end{align*}
Since $U$ is unitary the $\{\ket{\psi_i}=U\ket{i}\}$ are othornomal
and can be extended to form a complete basis of $\mathbb{C}^m$. $\quad\Box$

\subsection{Trace-preserving Quantum Operations}
\label{trace preservation subsection} The results of subsection
\ref{properties}, although extremely useful in quantum theory
(quantum information theory in particular), are in fact general
results on positive matrices and Completely Positive-preserving
linear maps. The same is true of subsection \ref{new properties},
and this is the reason why we have barely mentioned the unit trace
condition on density matrices so far. Yet in quantum theory the
states must have trace one (unless we start to consider the trace
as encoding some overall probability of occurrence), and quantum
operation must be Trace-preserving (so that they may always
occur). We now give an account of the main known results related
to these restrictions, augmented with some results stemming from
the state-operator correspondence.
\begin{Def} A linear map $\Omega : M_n(\mathbb{C})\rightarrow
M_m(\mathbb{C})$ is \emph{Trace-preserving} if and only if for all
$\rho$ in $M_n(\mathbb{C})$, $\trace(\Omega(\rho))=\trace(\rho)$.
\end{Def}
\begin{Def} The state $(1/d)Id_d\in\textrm{Herm}_{d}^+(\mathbb{C})$ is called the \emph{maximally
mixed} state of $\mathbb{C}^d$.\\
Moreover we say that $\$\in\textrm{Herm}_{mn}^+(\mathbb{C})$ is a
\emph{maximally entangled} state of $\mathbb{C}^m\otimes
\mathbb{C}^n$ if and only if $\$$ is pure and verifies either of
\begin{align*}
&(n \leq m)\quad\trace_1(\$)=Id_n\\
&(m \leq n)\quad\trace_2(\$)=Id_m
\end{align*}
depending on the integers $m$ and $n$ (if $m=n$ the two conditions
are indifferent).
\end{Def}
\begin{Lem}[Trace-preserving linear maps.]\label{trace preserving conditions}
A Completely positive-preserving linear map
$\whdoll:M_{n}(\mathbb{C})\rightarrow M_{m}(\mathbb{C})$ with
decomposition $\{\hat{A}_x\}$ is \emph{Trace-preserving} if and
only if one of the following six equivalent conditions is
satisfied:
\begin{align*}
&(i) \; \sum {\hat{A}}^{\dagger}_x \hat{A}_x=Id_n,  &(ii)\; \whdoll_{kk;jl}=\delta_{jl},\\
\textrm{In t}&\textrm{erms of }\wcdoll\textrm{ this is}\\
&(iii) \; \sum \check{A}_x \check{A}^{\dagger}_x=Id_n, &(iv) \; \wcdoll(Id_m)=Id_n,\\
\textrm{In t}&\textrm{erms of the state }{\$}\textrm{ this is}\\
&(v)\;\trace_1(\$)=Id_n,  &(vi)\; \$_{kj;kl}=\delta_{jl}.
\end{align*}
\end{Lem}
\emph{Proof.} We have $\whdoll(\rho)=\sum \hat{A}_x \rho
{\hat{A}}_x^{\dagger}$ or using components $\whdoll(\rho)_{i;k}=
\whdoll_{ik;jl}\rho_{jl}$, so that $\trace\big(\whdoll(\rho)\big)=
\trace\big(\sum {\hat{A}}^{\dagger}_x \hat{A}_x \rho \big)\equiv
\whdoll_{kk;jl}\rho_{jl}$. Thus $(i)$ and $(ii)$ follow
immediately. Using that $\check{A}= {\hat{A}}^t$ and
$\wcdoll_{jl;kk}=\wcdoll(Id_m)_{j;l}=\whdoll_{jl;kk}$
 from (\ref{shutranspose}),  we get $(iii)$ and $(iv)$. $(v)$ and
$(vi)$ follow from $(i)$ and $(ii)$ using (\ref{trace1}) and (\ref{shulonghat})
respectively.  $\quad\Box$ \\
Note that these conditions imply, but are not equivalent to,
$(1/n)\$$ having  unit trace. This is because `$\$$ has unit
trace' reads:
\begin{align*}
&\trace(\$)=\trace\big(\whdoll(Id_n)\big)=\trace\big(\wcdoll(Id_m)\big)=1 \\
\textrm{or}\;\;& \$_{kl;kl}=\whdoll_{kk;ll}=\wcdoll_{ll;kk}=1.
\end{align*}
Thus we have shown that Trace-preserving quantum operations
$\whdoll: M_n(\mathbb{C})\to M_m(\mathbb{C})$ go hand in hand with
unit trace states $(1/n)\$ \in \textrm{Herm}_{mn}^+(\mathbb{C})$
whose partial trace on the first subsystem yields the maximally
mixed state: $\trace_1\big((1/n)\$)=(1/n)Id_n$. We immediately
obtain the following, which is a generalization of a result in
\cite{Landau} and \cite{Verstraete}:
\begin{Lem}[Unitary maps.] \label{unitary operations}
Let $\whdoll: M_n(\mathbb{C})\to M_m(\mathbb{C})$ a Completely
Positive-preserving map. Then $\whdoll$ is \emph{isometric} (i.e
it can be written as $\whdoll:\rho \mapsto \hat{U}\rho
\hat{U}^{\dagger}$ with $\hat{U}^{\dagger}\hat{U}= Id_n$) if and
only if $n\leq m$ and the corresponding state $\$$ is
\emph{maximally entangled} (i.e. pure with $\trace_1(\$)=Id_n$).
Equivalently, in terms of indices, $\whdoll$ must verify
$\whdoll_{kk;jl}=\delta_{jl}$ and
\begin{equation*}
\sum_{jl}\trace\big({\whdoll(E_{jl})}^{\dagger}\whdoll(E_{jl})\big)=n^2
\end{equation*}
\end{Lem}
\begin{Rk}[Bistochastic maps.] \label{bistochastic maps}
$\whdoll$ is \emph{bistochastic}, i.e. it is Trace-preserving and
satisfies $\whdoll(Id_n)=Id_m$, if and only if the state $\$$
satisfies $\trace_1(\$)=Id_n$ and $\trace_2(\$)=Id_m$. Thus
bistochastic maps cannot be factorizable whenever $m\neq n$.
\end{Rk}
\emph{Proof.} The Lemma follows immediately from Lemma \ref{trace
preserving conditions} and Corollary \ref{operator purity}. The
remark follows from Lemma \ref{trace preserving conditions} and
the fact that $\trace_2(\$)=\whdoll(Id_m)$. $\quad \Box$

The set of states $\$\in \textrm{Herm}^+_{mn}(\mathbb{C})$
satisfying $\trace_1(\$)=Id_n$ is convex, hence its extremal
points correspond to extremal Trace-preserving quantum operations.
Recall that the extremal elements of a convex set $S$ are those
which cannot be written as sums of two distinct elements of $S$.
Extremal elements are important since they generate  $S$, and so
we now restate Choi's well-known theorem about extremal
Trace-preserving maps (without reproducing the proof).

\begin{Th}[Extremal Trace-preserving.]\label{extremal trace preserving}
Let $\whdoll:M_{n}(\mathbb{C})\rightarrow M_{m}(\mathbb{C})$ a
Trace-preserving Completely Positive-preserving linear map with
decomposition $\{\hat{A}_x\}$ and higher rank $r$ (i.e.
$r=rank(\$)$). Then $\whdoll$ is extremal in the set of
Trace-preserving Completely Positive-preserving maps if and only
if one of the following three
equivalent conditions is satisfied:\\
$(i)$ the span of the set $\{\hat{A}^{\dagger}_x \hat{A}_y\}$ in
$M_n(\mathbb{C})$  is $r^2$-dimensional;\\
$(ii)$ the span of the set $\{\check{A}_x \check{A}^{\dagger}_y\}$ in
$M_n(\mathbb{C})$  is $r^2$-dimensional;\\
$(iii)$ the span of the set $\{\trace_1(A_x A^{\dagger}_y)\}$ in
$M_n(\mathbb{C})$ is $r^2$-dimensional.
\end{Th}
Notice that this is a slightly different formulation from the one
given in \cite{Choi}, where the $\{\hat{A}^{\dagger}_x
\hat{A}_y\}$ have to form a linearly independent set. This implies
that the $\{\hat{A}_x\}$ are automatically linearly independent
themselves, and hence there must be $r$ of them. Since different
decompositions can give the same operation, we thought it better
to express the extremality conditions in terms of any
decomposition, and not just a minimal one.\\
\emph{Proof.} We just
prove the equivalence with Choi's formulation. Let
$\{\hat{V}_{\alpha}\}$ a minimal decomposition of $\whdoll$. Then
$\textrm{Span}(\{{A}_x\})=\textrm{Span}(\{V_{\alpha}\})$ since
both are equal to the support (the image space) of $\$$ (see
Corollary \ref{operator decomposition}); and trivially
$\textrm{Span}(\{\hat{A}_x\})=\textrm{Span}(\{\hat{V}_{\beta}\})$
implies $\textrm{Span}(\{\hat{A}^{\dagger}_x
\hat{A}_y\})=\textrm{Span}(\{\hat{V}_{\alpha}^{\dagger}
\hat{V}_{\beta}\})$.  $\quad \Box$ \\
\begin{Rk} An extreme map $\whdoll$ has higher rank $r\leq n$ since it
must satisfy $r^2\leq n^2$, but this condition is not sufficient.
\end{Rk}
\emph{Proof.} Suppose $U_1\neq U_2$ unitary and $\whdoll
:\rho\mapsto (1/2)U_1 \rho U_1^{\dagger}+(1/2)U_2 \rho
U_2^{\dagger}$. Clearly $U_1^{\dagger} U_1=U_2^{\dagger}
U_2=Id_n$, and thus this Trace-preserving Completely
Positive-preserving map cannot be extremal Trace-preserving. Yet
it has higher rank $2$ regardless of a choice for $n$.$\quad\Box$\\
By pushing the consequences of Choi's theorem further we obtain
the following original criteria for extremal Trace-preserving
linear maps:
\begin{Proposition}[Extremal Trace-preserving.]
Let $\whdoll:M_n(\mathbb{C}) \to M_m(\mathbb{C})$ a
Trace-preserving Completely Positive-preserving linear map of Choi
rank $r$ (i.e. $r=rank(\$)$) with $\$$ its corresponding state,
and ${\whdoll}^{{}^\dagger}:M_m(\mathbb{C}) \to M_n(\mathbb{C})$
its adjoint map (i.e.
${\whdoll}^{{}^\dagger}\equiv{\whdoll}^{{}^\dagger}_{jl;ik}$).
Then $\whdoll$ is extremal if and only if one of the
following equivalent conditions is satisfied:\\
$(i)$ The higher rank of ${\whdoll}^{{}^\dagger} \circ \whdoll$
is equal to $r^2$, \\
$(ii)$ $\$$ is such that the state in
$\textrm{Herm}^+_{n^2}(\mathbb{C})$ defined by
\begin{equation*}
\europ_{jj';ll'}=\$^*_{ij;kl}\$_{ij';kl'}
\end{equation*}
has rank $r^2$.
\end{Proposition}
\emph{Proof.} If $\whdoll$ has operator sum decomposition
$\{\hat{A}_y\}$ i.e. $\whdoll(\rho)=\sum_y \hat{A}_y \rho
\hat{A}_y^{\dagger}$, then we get that ${\whdoll}^{{}^\dagger}$
has decomposition $\{\hat{A}_x^{\dagger}\}$ i.e.
${\whdoll}^{{}^\dagger}(\sigma)= \sum_x \hat{A}_x^{{}^\dagger}
\rho \hat{A}_x$. This can be seen using
${\whdoll}^{{}^\dagger}_{jl;ik}\equiv \whdoll^*_{ik;jl}$ for
example. Thus ${\whdoll}^{{}^\dagger} \circ \whdoll$ has
decomposition $\{\hat{A}_x^{\dagger}\hat{A}_y\}$, and $(i)$, using
Corollary \ref{operator decomposition}, is equivalent to $(i)$
in Theorem \ref{extremal trace preserving}. \\
Next we restate $(i)$ using indices and Equation
(\ref{shulonghat}).
\begin{align*}
({\whdoll}^{{}^\dagger} \circ \whdoll)_{jl;j'l'}&={\whdoll}^{{}^\dagger}_{jl;ik}\whdoll_{ik;j'l'} \\
       &={\whdoll}^*_{ik;jl}\whdoll_{ik;j'l'}\\
       &=\$_{ij;kl}^*\$_{ij';kl'}\\
       &\equiv \wheurop_{jl;j'l'}= \europ_{jj';ll'}.
\end{align*}
Since ${\whdoll}^{{}^\dagger} \circ \whdoll$ is a Completely
Positive-preserving map from $M_n(\mathbb{C})$ to
$M_n(\mathbb{C})$, $\europ$ is in
$\textrm{Herm}^+_{n^2}(\mathbb{C})$
by Theorem \ref{operations as states}. We see that $(i)$ is equivalent to $(ii)$.  $\quad \Box$ \\
The relation between condition $(i)$ and $(ii)$ suggests that the
composition law on quantum operations could yield, through
Isomorphism \ref{isomorphism2}, an interesting structure upon
states. We pursue this idea in the following section.

\section{Induced geometrical structure}
\label{geometrical structure} The beginning of this section is
maybe aimed at a mathematically-minded reader. We investigate
simple algebraic and geometric properties stemming from the
operator state correspondence. These yield a nice group theoretic
description of totally entangled states of a bipartite system
(Proposition \ref{homomorphisms}), and a description of
Positive-preserving maps as dual to separable states (Theorem
\ref{Positive-preserving} restated). Proposition \ref{state as
measurement} however unravels a possible physical interpretation
of the correspondence.
\subsection{Composition laws}
We make use of some elementary facts about operators or
positive matrices to define new composition laws on the spaces of
operators or positive matrices.\\
First, the set of Completely Positive-preserving linear maps from
$M_n(\mathbb{C})$ into itself is stable under composition. This
induces the following semi-group structure for states (recall that
semi-group elements do not need to have an inverse):
\begin{Proposition} \label{semi-group1}
If $\,\$$ and $\europ$ are in $\textrm{Herm}_{n^2}^+(\mathbb{C})$,
then so is
\begin{equation}
\$ \diamond \europ \equiv (\$ \diamond
\europ)_{ij;kl}=\$_{ii';kk'}\europ_{i'j;k'l},
\end{equation}
where all the indices run from $1$ to $n$.
$(\textrm{Herm}_{n^2}^+(\mathbb{C}),\diamond )$ is a semi-group
with identity element the canonical maximally entangled state
$\ket{\beta}\bra{\beta}\equiv \delta_{ij}\delta_{kl}$.\\
The set of non-normalized pure states, the set of unentangled
states and the set of separable states (together with
$\ket{\beta}\bra{\beta}$), are sub-semi-groups of
$(\textrm{Herm}_{n^2}^+(\mathbb{C}),\diamond )$. More precisely,
\begin{align}
&(AA^{\dagger})\diamond (BB^{\dagger})=VV^{\dagger} \;\; where
\;\; \hat{V}=\hat{A}\hat{B} \label{pure to pure}\\
&(\mu_1\otimes \mu_2) \diamond (\sigma_1\otimes \sigma_2) =
\trace(\mu_2^t\sigma_1)\mu_1 \otimes \sigma_2 \nonumber
\end{align}
\end{Proposition}
\emph{Proof.} The composition law is just the transcription of
$(\whdoll\circ
\wheurop)_{ik;jl}=\whdoll_{ik;i'k'}\wheurop_{i'k';jl}$ using
(\ref{shulonghat}), and the identity element is clearly
$\delta_{ij}\delta_{kl}$. Next, the composition of two
factorizable operations is factorizable and trivially yields
(\ref{pure to pure}). Let $\europ=\sigma_1\otimes \sigma_2$ and
$\$=\mu_1\otimes \mu_2$ two unentangled states. We have using
Equation (\ref{dollarmment}):
\begin{align*}
\wheurop(\rho)&=\trace_2\big((Id\otimes
\rho^t)(\sigma_1\otimes \sigma_2)\big)=\big(\trace(\sigma_2^t\rho)\big)\sigma_1,\\
\textrm{hence} \;\; \whdoll\circ
\wheurop(\rho)&=\trace\big(\mu_2^t(\trace(\sigma_2^t\rho)\sigma_1)\big)\mu_1
\nonumber \\
 &= \trace(\mu_2^t\sigma_1)\trace(\sigma_2^t \rho) \mu_1 \nonumber
\end{align*}
and the last equation follows immediately.\\
Since the composition
law is bilinear, the space of separable states of
$\textrm{Herm}_{n^2}^+$, together with the identity
$\ket{\beta}\bra{\beta}$,  is also a sub-semi-group of
$(\textrm{Herm}_{n^2}^+(\mathbb{C}),\diamond )$.$\quad \Box$\\

It seems natural at this point to look for subgroups of
$(\textrm{Herm}_{n^2}^+(\mathbb{C}),\diamond )$. Clearly the
largest subgroup corresponds to the set of invertible quantum
operations $\whdoll$, of which it is difficult to give a physical
description in terms of the states $\$$: we just require
$\whdoll_{ik;jl}$ to be invertible. Since unentangled states yield
projections (as was illustrated in the proof above), they are not
in this group; yet mixtures of them (separable states) may well
yield invertible
operations.\\
\begin{Def} The positive definite matrices of $\textrm{Herm}_{d}^+(\mathbb{C})$
are sometimes called the \emph{totally mixed} states of $\mathbb{C}^d$.\\
Moreover we say that $\$\in\textrm{Herm}_{mn}^+(\mathbb{C})$ is a
\emph{totally entangled} state of $\mathbb{C}^m\otimes
\mathbb{C}^n$ if and only if $\$$ is pure and verifies either of
\begin{align*}
&(n \leq m)\quad\trace_1(\$)\;\textrm{is totally mixed}\\
&(m \leq n)\quad\trace_2(\$)\;\textrm{is totally mixed}
\end{align*}
depending on the integers $m$ and $n$ (if $m=n$ the two conditions
are indifferent).
\end{Def}
Let $GL_n(\mathbb{C})$ denote the group of invertible $n \times n$
complex matrices, $U(1)$ its (normal) subgroup of matrices of the
type $e^{i\theta}Id_n$, and $SU(n)$ the group of special unitary
$n\times n$ matrices, i.e. matrices $U$ satisfying $U^{\dagger}U=
UU^{\dagger}= Id_n$ and $\textrm{det}\,U=1$. We have the
following:
\begin{Proposition} \label{homomorphisms}The set of totally entangled pure states in
$\textrm{Herm}^+_{n^2}(\mathbb{C})$, equipped with the composition
law $\diamond$, is a group which is isomorphic  to the group
$GL_n(\mathbb{C}) / U(1)$. Its subset of maximally entangled
states is a subgroup isomorphic to $SU(n)$.
\end{Proposition}
\emph{Proof.} Let us denote by $T$ the set of totally entangled
(pure) states in $\textrm{Herm}^+_{n^2}(\mathbb{C})$. Note that
for any $\hat{A} \in M_n(\mathbb{C})$, $\hat{A}$ is invertible if
and only if $\hat{A}\hat{A}^{\dagger}$ is invertible, which by
(\ref{trace2}) is equivalent to $\trace_2(AA^{\dagger})$
invertible, in other words $AA^{\dagger}$ totally entangled. Thus
$T=\{AA^{\dagger}\;/\;\hat{A} \in GL_n(\mathbb{C}) \}$, and from
(\ref{pure to pure}), $(T, \diamond )$ is a group with identity
element $\ket{\beta}\bra{\beta}$.
\begin{align*}
\phi: GL_n(\mathbb{C}) & \to T \\
         \hat{A} \mapsto AA^{\dagger}
\end{align*}
is then trivially a group homomorphism, since
$\phi(\hat{A}\hat{B})= AA^{\dagger} \diamond BB^{\dagger}$ by (\ref{pure to pure}).
$\phi$ is clearly onto, and its kernel is $U(1)$.
Thus $GL_n(\mathbb{C}) / U(1)$ is isomorphic to $T$. \\
$\phi$ restricted to $U(n)$ maps onto the set of maximally
entangled states by Lemma \ref{unitary operations}, so that
$SU(n)=U(n)/U(1)$ is isomorphic to it. $\quad \Box$ \\

These results are useful when one seeks to parameterize certain
pure states of an $n^2$-dimensional system. The description of
pure states in $\textrm{Herm}_{n^2}(\mathbb{C})$ in terms of the
homogeneous space $SU(n^2)/SU(n^2-1)$ is well-known, but yields a
very complicated parameterization since one must mod out the
$SU(n^2-1)$. We have shown that we can in fact parameterize the
set of maximally entangled (pure) states of
$\textrm{Herm}_{n^2}(\mathbb{C})$ in terms of (the Euler angles
of) $SU(n)$, without having to mod out any redundancy. The
parameterization could have potential applications in the study of
entanglement, Bell states and EPR scenarios.

Next one can also define an original semi-group structure on the
set of Completely Positive-preserving maps by using an exotic
composition law (the Schur product $\bigtriangleup$) on the set of
states:
\begin{Proposition}
If $\whdoll$ and $\wheurop$ are Completely Positive-preserving
maps from $M_n(\mathbb{C})$ to $M_m(\mathbb{C})$, then so is
\begin{equation*}
\whdoll\bigtriangleup \wheurop\equiv (\whdoll\bigtriangleup
\wheurop)_{ik;jl}=\whdoll_{ik;jl}\wheurop_{ik;jl}
\end{equation*}
where the summation convention is suspended, and $i,k=1,\ldots,m$,
and $j,l=1,\ldots, n$. This composition law is obviously
commutative, and the set of factorizable operations is stable
under it.
\end{Proposition}
\emph{Proof.} This stems, via Theorem \ref{operations as states},
from the stability of the set of positive matrices under of the
Schur (or Hadamard) product \cite{matrices}. I.e. the fact that
the component-wise product of two positive matrices is a positive
matrix, when applied to $\$$ times $\europ$, induces the
corresponding result for $\whdoll$ times $\wheurop$. \\
We use the same symbol $\bigtriangleup$ to denote all
component-wise products of matrices. If $\whdoll$ and $\wheurop$
have decompositions $\{\hat{A}_x\}$ and $\{\hat{B}_y\}$
respectively, then $\whdoll \bigtriangleup \wheurop$ has
decomposition $\{\hat{A}_x \bigtriangleup \hat{B}_y \}$: this
implies the stability of factorizable operations. $\quad \Box$ \\

\subsection{Duality: states and functionals}
\label{duality}

When relating operators and states of a physical theory notions of
duality between vector spaces are often illuminating: operators
sometimes induce functionals on the space of states, which can in
turn be thought of as states.  In finite-dimensional Quantum
Mechanics, a given positive matrix can either represent a state or
a positive functional, and we can switch from one to the other easily.\\
So far we have equipped the algebra of complex $d\times d$
matrices, $M_d(\mathbb{C})$, with the complex-bilinear form:
$(\europ,\$)= \trace(\europ^{\dagger}\$)$. This non-degenerate
form naturally defines a canonical pairing of $M_d(\mathbb{C})$
with $\widetilde{M}_d(\mathbb{C})$, the space of linear
functionals on $M_d(\mathbb{C})$:
\begin{align*}
\widetilde{\phantom{U}}:\; M_d(\mathbb{C})&\longrightarrow \widetilde{M}_d(\mathbb{C})  \\
 \europ &\longmapsto [ \widetilde{\europ}: \$\mapsto \trace(\europ^{\dagger}\$) ]
\end{align*}
Since $\widetilde{\phantom{U}}$ is an (anti-linear) isomorphism,
any linear functional on $M_d(\mathbb{C})$ has a unique antecedent
by $\widetilde{\phantom{U}}$, thus is uniquely represented by an
element of $M_d(\mathbb{C})$. Let $\{E_{ij}\}_{1\leq i,j \leq d}$
a canonical basis of $M_d(\mathbb{C})$ and
$\{\widetilde{E}_{kl}\}_{1\leq k,l \leq d}$ its corresponding
peered basis, i.e.
$\widetilde{E}_{kl}(E_{ij})\equiv\trace(E_{kl}^{\dagger}E_{ij})=\delta_{ik}\delta_{jl}$.
Then the functional of $\europ$, namely $\widetilde{\europ}$, is
represented in the peered basis by $\europ^*$. Indeed,
$\europ^*_{kl}\widetilde{E}_{kl}(\$_{ij}E_{ij})= \europ^*_{kl}\$_{kl}=\widetilde{\europ}(\$)$. \\
When restricted to the real vector space of hermitian matrices
$\textrm{Herm}_d(\mathbb{C})$, $(\europ,\$)\mapsto
\trace(\europ\$)$ yields a real scalar product, and
$\widetilde{\textrm{Herm}}_d(\mathbb{C})$ is defined similarly. It
then becomes possible to define the dual (sometimes called polar)
of a subspace $\mathcal{S}$ of $\textrm{Herm}_d(\mathbb{C})$ as
follows:
\begin{equation}
\label{subspacedual} {\mathcal{S}}^{\star}\equiv \{
\widetilde{\sigma}\in \widetilde{\textrm{Herm}}_{mn}(\mathbb{C})
\; \;/ \;\; \forall \rho \in \mathcal{S}, \;\;
\widetilde{\sigma}(\rho)\geq 0 \; \}
\end{equation}
The convex cone of hermitian positive matrices
$\textrm{Herm}^+_d(\mathbb{C})$ is clearly self-dual under this
dual pairing:
\begin{align*}
\europ\in \textrm{Herm}^+_d(\mathbb{C}) &\Leftrightarrow  \forall
\$\in
\textrm{Herm}^+_d(\mathbb{C}), \;\; \trace(\europ\$) \geq 0 \\
&\Leftrightarrow  \forall \$\in
\textrm{Herm}^+_d(\mathbb{C}), \;\;\widetilde{\europ}(\$)\geq 0 \\
&\Leftrightarrow \widetilde{\europ} \in
{\textrm{Herm}^+_d(\mathbb{C})}^{\star}
\end{align*}
In the last line we have used the definition (\ref{subspacedual}).
Thus ${\herm}^{\star}=\widetilde{\textrm{Herm}}^+_d(\mathbb{C})$,
hence  the set of non-normalized states is isomorphic to that of
non-normalized linear probability distributions on states, i.e.
functionals which are positive on $\herm$. In this sense, if
$\europ$ is an element of $\herm$, then $\europ^*\equiv \europ^t$,
represents its dual element, or associated linear probability
distribution, and conversely. We shall now explain why this
picture is illuminating.

\subsubsection{Separable states and Positive-preserving maps}
Now call $\textrm{Herm}_{mn}^{S}(\mathbb{C})$ the set (convex
cone) of separable states of $\mathbb{C}^m \otimes \mathbb{C}^n$,
and define its dual space  by (\ref{subspacedual}):
\begin{align*}
&{\textrm{Herm}_{mn}^{S}(\mathbb{C})}^{\star} \equiv \\
&\{ \widetilde{\sigma}\in
\widetilde{\textrm{Herm}}_{mn}(\mathbb{C}) \;\; / \;\; \forall
\rho \in \textrm{Herm}_{mn}^{S}(\mathbb{C}), \;\;
\widetilde{\sigma}(\rho)\geq 0 \; \},
\end{align*}
This is a convex cone too. The geometrical meaning of Theorem
\ref{Positive-preserving} is now clear in this formalism:
\setcounter{Th}{1}
\begin{Th}[restatement.] \label{duality and separability}
A linear operation $\whdoll:M_n(\mathbb{C})\to M_m(\mathbb{C})$ is
Positive-preserving if and only if the linear functional of its
associated state $\$$, namely $\widetilde{\$}$, is in
${\textrm{Herm}_{mn}^{S}(\mathbb{C})}^{\star}$. In other words,
the convex cone of Positive-preserving maps is isomorphic to the
dual of the convex cone of separable states.
\end{Th}
Remember that inclusions are reversed by duality:
\begin{align*}
&\textrm{Herm}^S_{mn}(\mathbb{C}) \subsetneq
\textrm{Herm}_{mn}^+(\mathbb{C})\\
\Leftrightarrow &{\textrm{Herm}^+_{mn}(\mathbb{C})}^{\star}
\subsetneq {\textrm{Herm}_{mn}^S(\mathbb{C})}^{\star}.
\end{align*}
Since not all states are separable, this confirms the fact that
Positive-preserving maps are not necessarily Complete
Positive-preserving.

\begin{Rk}\label{partial transpose}
The set of $\$$ in $\textrm{Herm}_{mn}(\mathbb{C})$ such that
$\whdoll$ is Positive-preserving, i.e. such that $\widetilde{\$}$
belongs to ${\textrm{Herm}_{mn}^S(\mathbb{C})}^{\star}$, is stable
under the transposes $^{t_1}$ on $\mathbb{C}^m$  and $^{t_2}$ on
$\mathbb{C}^n$.
\end{Rk}
\emph{Proof:} For $\whdoll$ Positive-preserving, $\whdoll\circ
\,^{t_2}\equiv  \whx{\$^{t_2}}$ and $^{t_1} \circ \whdoll \equiv
\whx{\$^{t_1}}$ are Positive-preserving too. From this simple
observation we readily obtain that the set of the $\$$ is stable
under partial transpositions. $\quad \Box$\\
\begin{Rk}\label{peres criterion}
Remark \ref{partial transpose} is equivalent the Peres criterion
\cite{Peres} for separability, which states that the set of the
separable states $\textrm{Herm}_{mn}^S(\mathbb{C})$ is stable
under partial transposition.
\end{Rk}
\emph{Proof:} [Peres $\Rightarrow$ Remark \ref{partial transpose}]
If $\$$ is such that $\widetilde{\$}$ belongs to
${\textrm{Herm}_{mn}^S(\mathbb{C})}^{\star}$, then we have that
\begin{align*}
&\forall \europ \in
\textrm{Herm}_{mn}^S(\mathbb{C})\;\;\;\trace(\$\europ)\geq 0\\
\Rightarrow \;\;\;&\forall \europ \in
\textrm{Herm}_{mn}^S(\mathbb{C})\;\;\;\trace(\$\europ^{t_2})\geq
0\quad\textrm{by Peres}\\
\Rightarrow \;\;\;&\forall \europ \in
\textrm{Herm}_{mn}^S(\mathbb{C})\;\;\;\trace(\$^{t_2}\europ)\geq 0
\end{align*}
which means, by definition, that $\widetilde{\$^{t_2}}$ belongs to
${\textrm{Herm}_{mn}^S(\mathbb{C})}^{\star}$. The same applies with $^{t_1}$.\\
$[$Remark \ref{partial transpose} $\Rightarrow$ Peres $]$ Now let
$\$$ belong to ${\textrm{Herm}_{mn}^S(\mathbb{C})}$. Since
${\textrm{Herm}_{mn}^S(\mathbb{C})}$ is a closed convex set
containing $0$ we have, by the bipolar theorem (see for instance
\cite{bipolar}), that
${\textrm{Herm}_{mn}^S(\mathbb{C})}={\textrm{Herm}_{mn}^S(\mathbb{C})}^{\star\star}$.
Thus $\$$ belongs to
${\textrm{Herm}_{mn}^S(\mathbb{C})}^{\star\star}$, and so
\begin{align*}
&\forall \widetilde{\europ} \in
\textrm{Herm}_{mn}^S(\mathbb{C})^{\star}\;\;\;\trace(\$\europ)\geq 0\\
\Rightarrow \;\;\;&\forall \widetilde{\europ} \in
\textrm{Herm}_{mn}^S(\mathbb{C})^{\star}\;\;\;\trace(\$\europ^{t_2})\geq
0\quad\textrm{by Remark \ref{partial transpose}}\\
\Rightarrow \;\;\;&\forall \widetilde{\europ} \in
\textrm{Herm}_{mn}^S(\mathbb{C})^{\star}\;\;\;\trace(\$^{t_2}\europ)\geq
0
\end{align*}
which means, by definition, that $\$^{t_2}$ belongs to
${\textrm{Herm}_{mn}^S(\mathbb{C})}^{\star\star}={\textrm{Herm}_{mn}^S(\mathbb{C})}$.
The same applies with $^{t_1}$: we have recovered Peres' criterion. $\quad \Box$\\
That the Peres' criterion corresponds to the simple fact that
$\whdoll$ Positive-preserving implies $\whdoll\circ^t$
Positive-preserving is a somewhat striking fact. This insight may
well help to build tighter criterions: recently the Horodeckis
\cite{Horodeckis} have been following this line of thought.

\subsubsection{Physical interpretation of formulae}
When attempting to characterize separability the notions of
duality seem to play a simplifying role, as they help to clarify
the correspondence induced by Isomorphism $2$. Thus one may wonder
if these concepts could facilitate the interpretation of other
results in this article. We now give a formulation of quantum
operations $\whdoll$ in terms of single operations on their
corresponding state $\$$.
\begin{Proposition} \label{state as measurement}
Let $\$$ represent a non-normalized quantum state of a bipartite
system $\mathcal{H}_A \otimes \mathcal{H}_B = \mathbb{C}^m\otimes
\mathbb{C}^n$ shared by Alice and Bob. Suppose Bob performs on
$\$$ a local generalized measurement $\{ Id_m \otimes M_n^{(x)}
\}_x$. Call $Id_m\otimes M$ the element whose outcome occurs and
let $\rho_B\equiv (M^{\dagger}M)^t
\in \textrm{Herm}_n^+(\mathbb{C})$. \\
Then the unrescaled post-measurement state as viewed by Alice  is
precisely $\whdoll(\rho_B)$. Thus the effect of any quantum
operation $\whdoll$ can be viewed as the trace out of a particular
local single operation on its corresponding state $\$$.
\end{Proposition}
\emph{Proof:} The unrescaled post-measurement state is simply
$\$^M=(Id_m \otimes M)\$(Id_m \otimes M^{\dagger})$.
Using (\ref{dollarmment}) this yields for Alice the state:
\begin{align*}
\trace_2(\$^M) & = \trace_2\big((Id_m\otimes M)\$(Id_m\otimes
M^{\dagger})\big) \\
& = \whdoll\big((M^{\dagger}M)^t\big) \\
& = \whdoll(\rho_B) \quad \Box
\end{align*}
The fact that there is a transpose corroborates the idea of
duality. Indeed, first $M^{\dagger}M$ is thought of as defining a
functional $\sigma \mapsto \trace(M^{\dagger}M\sigma)$, but then
as we think of a quantum operation as acting on states we act upon
its transpose. The map $M^{\dagger}M \mapsto
\whdoll\big((M^{\dagger}M)^t\big)$, though it is
Positive-preserving,  is not Completely Positive-preserving since
it can be written as $\whdoll \circ \,^{t}$. However the same map
defined from states to states, i.e.
$(M^{\dagger}M)^t\mapsto\whdoll((M^{\dagger}M)^t)$, is Completely
Positive-preserving. Proposition \ref{state as measurement}
suggests that quantum states in $\textrm{Herm}_{mn}^+(\mathbb{C})$
inherently defines a quantum operation between their two
subsystems.

\begin{table}[h]\caption{Summary}\label{summary}
\begin{tabular}{|l|l|}\hline\hline
\emph{Matrix} $\$$ &\emph{Linear operator} $\whdoll$\\
\hline\hline Hermitian              &Hermitian-preserving \\
\hline Dual to separable            &Positive-preserving \\
\hline Positive                     &Completely
Positive-preserving\\

\hline\hline \emph{Particular state} $\$$ &\emph{Particular quantum operation} $\whdoll$\\
\hline\hline Pure                   &Factorizable\\
\hline Unentangled $\sigma_1\otimes\sigma_2$& Projection
$\rho\mapsto\big(\trace_2(\sigma_2^t\rho)\big)\sigma_1$\\
\hline Separable                    &Sum of projections\\
                                    &Dual to Positive-preserving \\
\hline $\trace_1(\$)=Id$            &Trace-preserving\\
\hline $\trace_1(\$)=Id$ and $\trace_2(\$)=Id$
&Bistochastic\\

\hline\hline\emph{Particular ket} $A$ &\emph{Particular
evolution matrix} $\hat{A}$\\
\hline\hline Maximally entangled    &Unitary\\
\hline Totally entangled            &Invertible\\
\hline $\sum_i\ket{i}\ket{i}$       &$Id$\\
\hline $\sum_i\lambda_i\ket{i}\ket{i}$
&$Diag\{\lambda_i\}$\\\hline
$\sum_i\lambda_i\ket{\psi_i}\ket{\psi_i^*}$&\\
with $\forall i,\lambda_i\in\mathbb{R}$     &Hermitian\\
with $\forall i,\lambda_i\in\mathbb{R}^+$   &Positive\\

\hline\hline
\emph{Theorems on states} &\emph{Theorems on quantum operations}\\
\hline\hline
Spectral decomposition,     &Operator Sum decomposition,\\
Unitary degree of freedom   &Unitary degree of freedom\\ \hline
Purification &$\whdoll(\rho)=\trace_1(U\rho U^{\dagger})$\\
\hline Bipartite decompositions: &Matrix decompositions: \\
Schmidt                     &Polar\\
One-sided triangular        &QR\\
Two-sided triangular        &Schur's triangularization \\
\hline Purity condition        &Factorizability condition\\

\hline\hline \emph{Formulae on states}   &\emph{Formulae on
quantum operations}\\
\hline\hline
$\trace_{1/2}(AB^{\dagger})$&$=(\hat{B}^{\dagger}\hat{A})^{t}$/$\hat{A}\hat{B}^{\dagger}$\\
\hline $\trace_2\big((\kappa\otimes\rho^t)\$(\tau\otimes\sigma^t)\big)$&$=\kappa\whdoll(\rho\sigma)\tau$\\
\hline $\trace_2(Id\otimes\rho)\$(Id\otimes\rho^{\dagger}))$&$=\whdoll((\rho^{\dagger}\rho)^t)$\\
\hline $\trace((\sigma\otimes\rho^t)\$)$&$=\trace(\sigma\whdoll(\rho))$\\
\hline
$\trace(\europ^{\dagger}\$)$&=$\sum\trace({\wheurop(E_{jl})}^{\dagger}\,\whdoll(E_{jl}))$\\
\hline\hline
\end{tabular}
\end{table}

\section{Summary and concluding remarks}
\label{conclusion} In this article we make several new
contributions, some technical, others more geometrical.\\
Amongst the technical results we provide two triangular
decompositions for pure states of a bipartite system, i.e. local
changes of basis so that vectors in $\mathbb{C}^m \otimes
\mathbb{C}^n$ may be written with triangular coefficients only. We
also give two original algebraic tests on Completely
Positive-preserving maps: one regarding the factorizability or
single operator decomposition, the other testing extremality in
the set of Trace-preserving operations. These are particularly
interesting in the sense that they do not depend on the operator sum
decompositions of these maps. The formulae in Proposition \ref{multiplication}
should yield simplifications in optimization of fidelities of quantum
operations as encountered for instance in quantum cryptographic problems. \\
On the more geometrical side we endow
$\textrm{Herm}_{n^2}^+(\mathbb{C})$ with a semi-group structure
stemming from the composition law on quantum operations. This in
turn yields a group isomorphism between totally entangled (pure)
states and $GL_n(\mathbb{C})/U(1)$, and maximally entangled (pure)
states and $SU(n)$. This result sheds light on the geometry of
entangled states as it suggests, for future work, simple
parameterizations and bi-invariant metrics on the corresponding
(group-)submanifolds of the set of pure states in
$\textrm{Herm}_{n^2}^+(\mathbb{C})$. In addition we show that the
set of quantum operations is stable under component-wise product.\\
These contributions are interesting enough by themselves, but
perhaps the most significant achievement of this article is to
demonstrate the central, transversal role of the state-operator
isomorphism as formalized in Isomorphism $2$ and justified by
Theorem \ref{operations as states}. We have shown that virtually
all the main results regarding states/operators can be elegantly
brought as corollaries of their operator/state analogue, which
makes this correspondence one of the most fruitful linear
algebraic tool in the surroundings of quantum theory (see table
\ref{summary} for summary). Even for more specialist issues of
quantum information theory we find that the isomorphism has a role
to play, as was
illustrated by the problem of characterizing separable states. \\
On this occasion we introduced notions of duality, which serve
both to facilitate the interpretation of the state-operator
correspondence and its related formulae, and to understand the
underlying geometry from a slightly more abstract point of view.
The formulae themselves should have numerous applications in
quantum information theory, and maybe (as suggested in Proposition
\ref{state as measurement}) provide a novel interpretation of
states versus operations in open systems.

\section*{Acknowlegments}
C.E.P would like to thank Gary Gibbons for motivating discussions,
EPSRC, the DAMTP, and the Cambridge European and Isaac Newton
Trusts for financial support. P.J.A  would like to thank Anuj
Dawar for his patient listening, EPSRC, Marconi, the Cambridge
European and Isaac Newton Trusts for financial support.


\begin{thebibliography}{99}
\bibitem{Jamio} A. Jamiolkowski \emph{Linear transformations which
preserve trace and positive semidefinite operators}, Rep. Mod.
Phys., $\mathbf{3}$, 275-278, (1972).
\bibitem{Choi} M.D. Choi, \emph{Completely Positive linear maps on
complex matrices}, Lin. Alg. Appl., $\mathbf{10}$, 285-290, (1975).
\bibitem{Kraus1} K. Kraus, \emph{General state changes in quantum
  theory}, Annals of Physics, $\mathbf{64}$, 311-315, (1971). 
\bibitem{Kraus} K. Kraus, \emph{States Effects and Operators:
Fundamental Notions of Quantum Theory}, Springer Verlag, (1983).
\bibitem{Landau} L.J. Landau, R.F. Streater, \emph{On Birkhoff's
Theorem for Doubly Stochastic Completely Positive Maps of Matrix
Algebras}, Lin. Alg. Appl. $\mathbf{193}$, 107, (1993).
\bibitem{Su} E.C.G Sudarshan, P. M. Mathews, J. Rau, \emph{Stochastic Dynamics of Quantum-Mechanical Systems}, Phys. Rev. $\mathbf{121}$, 920-924, (1961).
\bibitem{Shu} E.C.G Sudarshan, \emph{Quantum Measurements and
Dynamical maps}, in \emph{From SU(3) to Gravity}, Ed. E. Gotsman,
G. Tauber, Cambridge University Press, (1986).\\
E.C.G. Sudarshan, A. Shaji, \emph{Structure and
Paramatrization of Generic Stochastic Maps of Density Matrices},
J. Phys. A, $\mathbf{36}$, 5073-5081, (2003).
\bibitem{decoys} P.J. Arrighi, \emph{Quantum Decoys},
  arXiv: quant-ph/0308050. 
\bibitem{matrices} R.A. Horn, C.R. Johnson, \emph{Matrix Analysis},
Cambridge University Press, (1985).
\bibitem{Horodeckis} M. Horodecki, P. Horodecki, R. Horodecki,
\emph{Separability of mixed quantum states: necessary and
sufficient conditions} Phys. Lett., $\mathbf{223}$, 1, (1996).
\bibitem{Nielsen} M.A. Nielsen, I.L. Chuang, \emph{Quantum
Computation and Quantum Information}, Cambridge University Press,
(2000).
\bibitem{Peres} A. Peres, \emph{Separability criterion for density
matrices}, Phys. Rev. Lett., $\mathbf{77}$, 1413-1415, (1996).
\bibitem{depillis} J. de Pillis, \emph{Linear transformations which
preserve hermitian and positive semidefinite operators}, Pacific
J. Math. $\mathbf{23}$, 129-137, (1967).
\bibitem{Hawking} S.W. Hawking, \emph{Breakdown of predictability
in gravitational collapse.}, Phys. Rev. D. (3), $\mathbf{14}$, no.
10, 2460-2473,(1976)
\bibitem{Verstraete} F. Verstraete, H. Verschelde, \emph{On quantum
Channels.}, Internal Report 02-176, ESAT-SISTA, K.U.Leuven,
(2002).
\bibitem{bipolar} C. Wagschal \emph{Topologie et analyse fonctionnelle}
Hermann Editeur des Sciences et des Arts, Méthodes N° 27669,
(1995).
\end{thebibliography}
\end{document}